\documentclass[useAMS,usenatbib,usegraphicx]{mn2e}

\newcommand{\arcdeg}{^\circ}
\usepackage{rotating}
\usepackage{multirow}
\usepackage{longtable}
\usepackage{amsmath}
\usepackage{amssymb}
\title[Optical line ratios in IC\,1805]{Diagnostic line ratios in the IC\,1805 optical gas complex}

\author[Lagrois, Joncas, \& Drissen]{Dominic Lagrois$^{1,2}$\thanks{E-mail: dominic.lagrois.1@ulaval.ca (D. L.)}, Gilles Joncas$^{1,2}$, and Laurent Drissen$^{1,2}$\\
$^{1}$ D\'epartement de physique, de g\'enie physique et d'optique, Universit\'e Laval, Qu\'ebec, QC, CANADA G1V 0A6\\
$^{2}$ Centre de Recherche en Astrophysique du Qu\'ebec, CANADA\\}

\date{Released 2011 Xxxxx XX}

\pagerange{\pageref{firstpage}--\pageref{lastpage}} \pubyear{2011}

\begin{document}

\maketitle

\label{firstpage}

\begin{abstract}
Large H\,\textsc{ii} regions, with angular dimensions exceeding 10 pc, usually enclose numerous massive O-stars. Stellar winds from such stars are expected to play a sizeable role in the dynamical, morphological and chemical evolution of the targeted nebula. Kinematically, stellar winds remain hardly observable i.e., the typical expansion velocities of wind-blown bubbles being often confused with other dynamical processes also regularly found H\,\textsc{ii} regions. However, supersonic shock waves, developed by stellar winds, should favor shock excitation and leave a well-defined spectral signature in the ionized nebular content. In this work, the presence of stellar winds, observed through shock excitation, is investigated in the brightest portions of the Galactic IC\,1805 nebula, a giant H\,\textsc{ii} region encompassing at least 10 O-stars from main-sequence O9 to giant and supergiant O4. The use of the imaging Fourier transform spectrometer SpIOMM enabled the simultaneous acquisition of the spectral information associated to the H{$\alpha$}\,$\lambda$6563 $\mbox{\AA}$, $[$N\,\textsc{ii}$]$\,$\lambda$$\lambda$6548, 6584 $\mbox{\AA}$, and $[$S\,\textsc{ii}$]$\,$\lambda$$\lambda$6716, 6731 $\mbox{\AA}$ ionic lines. Diagnostic diagrams, first introduced by Sabbadin and collaborators, were used to circumscribe portions of the nebula likely subject to shock excitation from other areas dominated by photoionization. The gas compression, expected from supersonic shocks, is investigated by comparing the pre- and post-shocked material's densities computed from the $\frac{[\textnormal{S}\,\textsc{ii}]\,\lambda6716}{[\textnormal{S}\,\textsc{ii}]\,\lambda6731}$ line ratio. The typical $\frac{[\textnormal{N}\,\textsc{ii}]\,\lambda6584}{[\textnormal{N}\,\textsc{ii}]\,\lambda6548}$ line ratio slightly exceeds the theoretical value of 3 expected in low-density regimes. To explain such behavior, a scenario based on collisional de-excitations affecting the $[$N\,\textsc{ii}$]$\,$\lambda$6548 $\mbox{\AA}$ line is proposed.
\end{abstract}

\begin{keywords}
techniques: spectroscopic --- ISM: H\,{\scriptsize{II}} regions --- ISM: lines and bands --- ISM: individual: IC\,1805
\end{keywords}

\section{Introduction}

The presence of ionized material in the interstellar medium (hereafter, ISM) can be attributed to two distinctive mechanisms. First, photoionization of the surrounding neutral gas by the strong, energetic ultra-violet (UV) flux of nearby massive stars is largely responsible for the detection of the standard hydrogen Balmer series, typically used for the morphological and kinematical description of H\,\textsc{ii} regions. Secondly, stellar winds with high terminal velocities and violent supernova blasts are commonly associated with the propagation of transonic and supersonic shock waves in the surrounding medium. The important increase of the post-shocked gas' temperature favors its ionization through shock excitation.

Pioneering work by \citet{Sab1977} has compared specific flux ratios for a variety of ionic transitions in the ISM optical gas. This allowed the authors to approximately separate standard H\,\textsc{ii} regions and planetary nebulae (hereafter, PNe) mostly governed by photoionization from shock-dominated supernova remnants (hereafter, SNRs). These diagnostic diagrams were often used, in literature, to classify large amount of ionized targets in large-scale objects, for example more-or-less distant galaxies (e.g., \citealt{Mag2003,Rie2006}). This has led to emission-line ratio plots in which each object is usually statistically represented by a single point. Obviously, intrinsic variations, within a given object, can be investigated by targeting Galactic objects individually (e.g., \citealt{Phi1998,Phi1999,Phi2010}). This allows to spatially resolve much smaller ($\ll$\,0.1 pc) structures and artifacts characterized by peculiar line ratios that are, otherwise, unrevealed or statistically negligible in observations using poorer angular resolutions. The investigation of close ionized objects has already revealed that photoionization and shock excitation can both be found in individual regions.

Kinematically speaking, standard H\,\textsc{ii} regions encompassing massive stars earlier than O6 have shown little indication of a strong impact on the surrounding gas attributed to stellar winds, referred earlier as a potential source for shock excitation in nebulae. This remains usually true even when using observations with high spectral resolutions that allow to measure velocity fluctuations down to a few km s$^{-1}$. The main reason for this resides in the double-shock model said to accurately describe the dynamical evolution of wind-blown bubbles \citep{Wea1977}. The reverse shock quickly (i.e., often within less than a spatial element of resolution) converts high-velocity ($>$\,1\,000 km s$^{-1}$) terminal winds into low-velocity, high-temperature gas. The thermal energy of this hot, pressurized material then initiates and fuels the expansion of the dense shell of post-shock ISM material found at much greater distances with respect to the central star. The typical expansion velocity of the shocked shell is usually a few km s$^{-1}$ above 10 km s$^{-1}$, roughly the speed of sound for warm ($\sim$10\,000 K) H$^{+}$ gas. Unfortunately, in complex tridimensional geometries, this kind of velocities could be easily confused with standard accelerated outflows in H\,\textsc{ii} regions, such as Champagne \citep{Ten1979,Bod1979} or photoevaporated \citep{Art2006,Mel2006} flows, turbulent motions \citep{Jon1984,Ars1988,Lag2011} or fluid instabilities.  This work mainly explores the possibility of detecting shock waves associated, in particular, to stellar winds using the emissive properties of the ionized gas rather than a more typical, and not always successful, approach based on the information retrieved from radial velocities and non-thermal line widths.

The IC\,1805 nebula is located in the Perseus arm of our Galaxy. The most massive stars of the Melotte 15 star cluster are currently responsible for the energetic support of the large H\,\textsc{ii} region. The south-central portion of the nebula (see Figure 1), in the direct vicinity of the star cluster, is gas-rich and will be used in this work to fill the \citet{Sab1977} diagrams. Our goals are to (1) obtain reliable, high signal-to-noise ratios (hereafter, S/N), spectra of the optical emission in the brightest portions of the IC\,1805 star-forming complex, (2) provide the corresponding series of line-ratio diagnostic diagrams, and (3) investigate the impact of photoionization and shock excitation in the targeted gas volume.

We present, in $\S$~2, the Galactic H\,\textsc{ii} region IC\,1805 and its associated star cluster, Melotte 15. Information related to our spectrometric observations and methods used for data reduction are detailed in $\S$~3. Results of our study and diagnostic diagrams are provided in $\S$~4. Interpretation and discussion follow in $\S$~5. Summarized results and general conclusions will finally be provided in $\S$~6.

\section{The IC\,1805 gas complex}

Introduced by \citet{Wes1958} as an irregular nebula of medium brightness, the Galactic H\,\textsc{ii} region IC\,1805 is often referred as the Heart nebula due to the heart-shaped morphology of its southern hemisphere (60$\arcdeg$28$\arcmin$\,$\leqslant$\,$\delta_{2000}$\,$\leqslant$\,62$\arcdeg$10$\arcmin$). It is suggested that outflows from the OB association Melotte 15 ($\alpha_{2000}$\,=\,02$^{\textnormal{h}}$32$^{\textnormal{m}}$45$^{\textnormal{s}}$, $\delta_{2000}$\,=\,61$\arcdeg$26$\arcmin$40$\arcsec$) are responsible for the actual expansion of the large H\,\textsc{ii} region/superbubble \citep{Bas1999}. Proper motion investigation has allowed the identification of 126 stars intrinsic to the star cluster \citep{Shi1999}. The authors confirmed, through spectroscopic observations, the presence of numerous massive stars; about forty from spectral types O4 to B2 (including 10 O-type stars) have been identified with a high probability of membership. An important proportion of these massive stars are still found on the main-sequence branch indicating a very young cluster age evaluated at 2.5 Myr \citep{Llo1982}. The dynamical age of the IC\,1805 nebula being estimated at 14 Myr \citep{Lag2009b}, it appears that the large H\,\textsc{ii} region was formed and shaped by the mechanical deposit, in the ISM, of energy (i.e., photon flux, stellar winds, supernovae) attributed to a succession of different star clusters, Melotte 15 being the most recently formed. No non-thermal emission being actually detected in the large star-forming complex, this suggests that the impact of old supernovae could be negligible in the nebula's current dynamics. UBV photometry, properly corrected for absorption, puts the Melotte 15 complex at an heliocentric distance of 2.35 kpc \citep{Mas1995}.

\citet{Lag2009a,Lag2009b} provide a high spectral-resolution ($<$\,15 km s$^{-1}$) kinematical investigation of the H$^{+}$ content on an appreciable fraction of the IC\,1805 ionized extent. The northernmost portions (63$\arcdeg$20$\arcmin$\,$\leqslant$\,$\delta_{2000}$\,$\leqslant$\,66$\arcdeg$50$\arcmin$) of the large ovoid bubble have shown kinematical evidences for high-latitude gas venting and the early development of a Galactic chimney. It appears that leaking UV photons, emanating from the Melotte 15 star cluster and unabsorbed by the denser low-declination material, may participate in the sustainment of the high-temperature gas found in the Galactic halo. In the southern half of the nebula, schematically represented in Figure 1, the last residual fragments of an old giant molecular cloud appear to be eroded by the UV flux and stellar winds of the nearby massive stars. As a result, series of criss-crossing accelerated flows are detected in agreement with the Champagne phase (e.g., \citealt{Ten1979,Bod1979,Ten1981,Ten1982}).

\begin{figure*}
  \begin{center}
    \leavevmode
    \scalebox{0.8}{\includegraphics[scale=1.10]{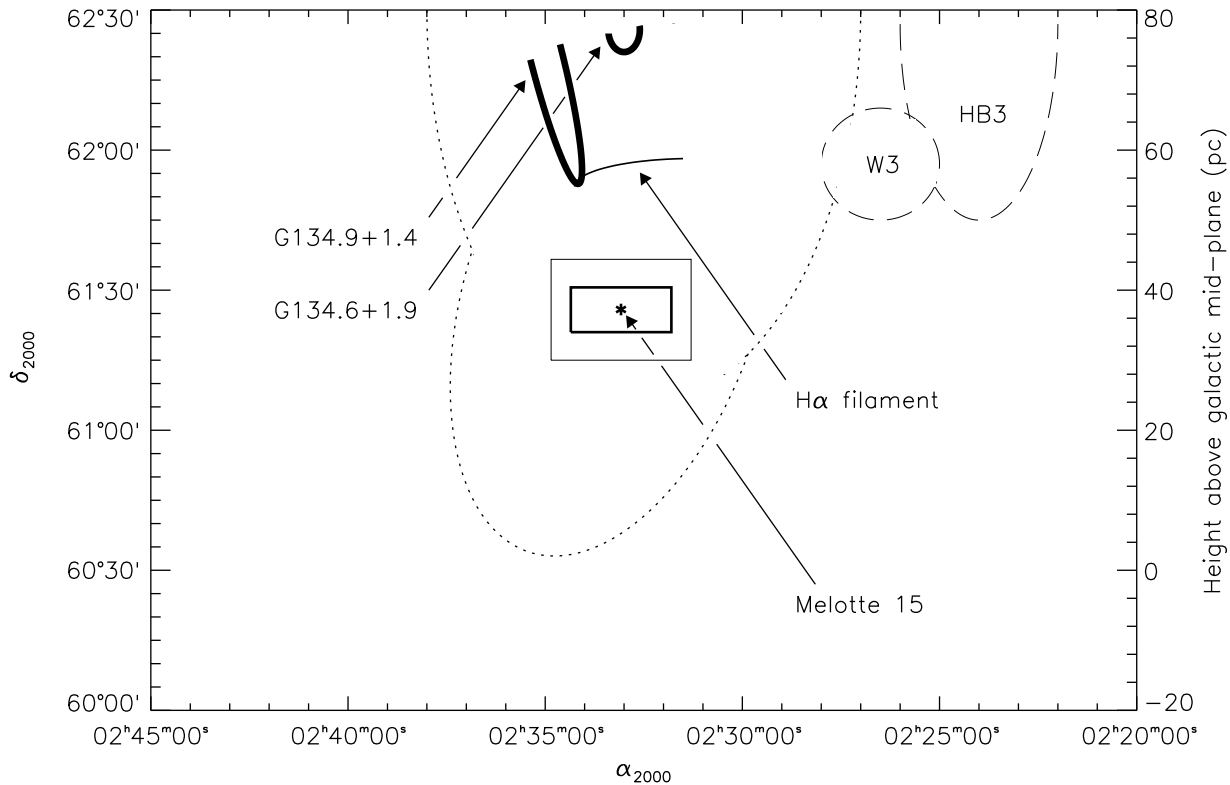}}
  \end{center}
  \caption{Schematic diagram of the southern portions of the IC\,1805 nebula (circumscribed by the dotted line). The regular and thick lines respectively traces out the H$^{+}$ and CO (2-1) structures (see Figure 1 of \citet{Lag2009a}). The thin-line box approximately delineates the gas-rich region in the vicinity of the Melotte 15 star cluster. The thick-line box approximately indicates the extended FOV of our spectrometric observations. Also included to the diagram, the nearby H\,\textsc{ii} region W3 and the supernova remnant HB3 (long-dashed lines).}
\end{figure*}

Ionization fronts, stellar winds, and Champagne shocks are therefore expected to have developed in the vicinity of the Melotte 15 star cluster. This could suggest that shock excitation may be responsible for the presence of a certain fraction of the observed ionized material although a well-defined kinematical signature attributed to shock waves has not been formally identified in our previous study. This will be largely addressed in the following sections.

\section{Observations and data reduction}

\begin{table*}
 \centering
 \begin{center}
 \caption{Observational Parameters at Data Acquisition}
 \vskip 0.3cm
 \begin{tabular}{c c c c c}
 \hline
 \hline
 & & & \multicolumn{2}{c}{Field} \\
 \multicolumn{2}{c}{Observational Parameters} & & East & West \\
 \cline{1-2} \cline{4-5}
 \multirow{2}*{Optical center} & $\alpha_{2000}$ & & 02$^{\textnormal{h}}$33$^{\textnormal{m}}$55$^{\textnormal{s}}$ & 02$^{\textnormal{h}}$32$^{\textnormal{m}}$35$^{\textnormal{s}}$ \\
                               & $\delta_{2000}$ & & 61$\arcdeg$27$\arcmin$30$\arcsec$ & 61$\arcdeg$27$\arcmin$30$\arcsec$ \\
 Seeing at acquisition & ($\arcsec$) & & 2.5 & 2.8 \\
 \multirow{2}*{Spatial resolution} & ($\arcsec$ pixel$^{-1}$) & & 2.2 & 2.2 \\
                                   & (pc pixel$^{-1}$) & & 0.026 & 0.026 \\
 Spectral Range & ($\mbox{\AA}$$-$$\mbox{\AA}$) & & 6480$-$6820 & 6480$-$6820 \\
 \multirow{2}*{Spectral resolution} & ($\mbox{\AA}$) & & 3 & 3 \\
                                    & (km s$^{-1}$) & & 150 & 150 \\
 Exposure time per channel & (s) & & 32 & 35 \\
 Number of channels & & & 249 & 249 \\ 
 \hline
 \end{tabular}
 \end{center}
\end{table*}

Observations of the IC\,1805 optical gas complex were performed during the nights of 2008 September 24-25, 25-26 using the Ritchey-Chr\'etien 1.6m telescope of the Observatoire du Mont-M\'egantic (OMM). The data were gathered using the imaging Fourier transform spectrometer SpIOMM (Spectrom\`etre Imageur de l'Observatoire du Mont M\'egantic). Technical details regarding SpIOMM are provided in \citet{Ber2008}, \citet{Dri2008} and references therein. 

For each pixel of the detector, the instrument has the capacity to obtain the emission spectrum of the corresponding nebular-gas column in selected bandwidths of the optical regime between 3500 $\mbox{\AA}$ and 9000 $\mbox{\AA}$. Prior to entering the observation mode, the spectral resolution is fixed by the observer between R\,=\,1 and 25\,000. 

For this work, the use of an interference filter enabled the acquisition of the spectral information in the red portion of the electromagnetic spectrum between 6480 $\mbox{\AA}$ and 6820 $\mbox{\AA}$. This allowed to obtain, simultaneously, emission spectra displaying the H{$\alpha$}\,$\lambda$6563 $\mbox{\AA}$, $[$N\,\textsc{ii}$]$\,$\lambda$$\lambda$6548, 6584 $\mbox{\AA}$, and $[$S\,\textsc{ii}$]$\,$\lambda$$\lambda$6716, 6731 $\mbox{\AA}$ ionic transitions. A spectral resolution of R\,=\,2100 was judged appropriate corresponding to a resolution of roughly 3 $\mbox{\AA}$ or 150 km s$^{-1}$ centered on the H{$\alpha$} rest frequency. The field-of-view (FOV) is roughly 12$\arcmin$\,$\times$\,12$\arcmin$ while data were spatially binned 2\,$\times$\,2 during acquisition to reduce readout times.

\begin{figure*}
  \begin{center}
    \leavevmode
    \scalebox{0.8}{\includegraphics[scale=1.10]{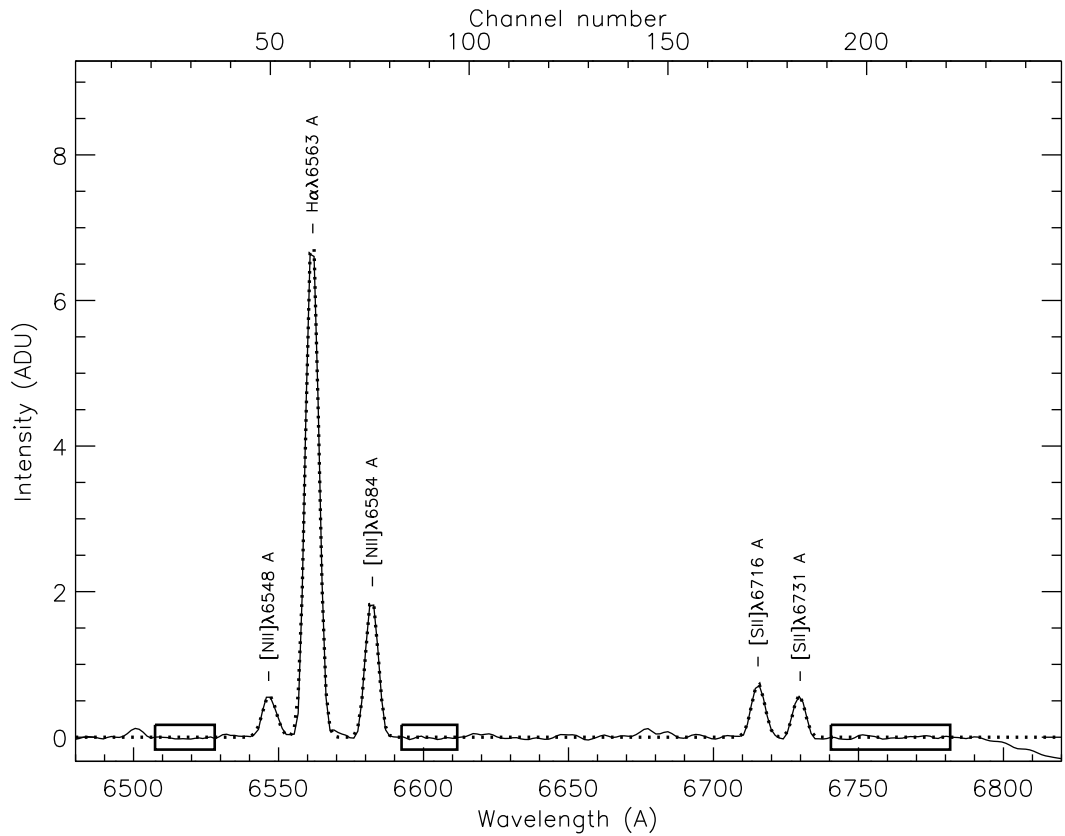}}
  \end{center}
  \caption{High-quality emission spectrum recovered from our spectrometric observations of the brightest parts of the IC\,1805 gas complex. The raw data are displayed as the regular line. Gaussian fits are represented as the dotted line. The boxes indicates the inter-line channels where the noise level was estimated. All five ionic transitions, targeted in this work, are listed. The S/N varies from 24 to 298 from $[$N\,\textsc{ii}$]$\,$\lambda$6548 $\mbox{\AA}$ to H{$\alpha$}\,$\lambda$6563 $\mbox{\AA}$.}
\end{figure*}

The resulting raw data cube is formed, for each pixel, of a discrete interferogram. Following classical data reduction for CCD observations (i.e., bias subtraction and flatfield correction), all panchromatic images were realigned to correct for guiding errors. According to the photometry of stars located in gas-depleted zones, all images were also corrected for variations of the sky transparency occurring during data acquisition. Each interferogram is then Fourier transformed and calibrated in velocities using a He-Ne laser ($\lambda$6328 $\mbox{\AA}$) as frequency of reference. The data were again spatially binned 2\,$\times$\,2 during reduction in order to account for the 2$\buildrel\arcsec\over.$5 seeing measured at the telescope. The resulting spatial resolution is of 2$\buildrel\arcsec\over.$2 pixel$^{-1}$ for 325\,$\times$\,335 pixel$^{2}$. Using the adopted distance to Melotte 15 (see $\S$~2), this corresponds to an angular resolution of 0.026 pc pixel$^{-1}$ as seen on the plane of the sky.

Two cubes were obtained in the vicinity of the Melotte 15 star cluster. The observational parameters for both cubes, labeled East and West, are summarized in Table 1. The heliocentric correction being very similar for both cubes, the two were mosaicked to a new extended FOV of 21$\buildrel\arcmin\over.$5\,$\times$\,12$\buildrel\arcmin\over.$5 (see Figure 1) and over 200\,000 emission spectra. Following subtraction of the continuum, a multi-component Gaussian fit procedure, written in \textsc{idl}, was applied to the mosaicked cube. The fitting procedure returns, for each component, the peak intensity in Analog-to-Digital Units (ADU), the line centroid (or velocity), and the line dispersion ($\sigma$\,$\equiv$\,$\frac{\textnormal{FWHM}}{2.354}$). Since this work is mostly based on line ratios, we followed the recommendation of \citet{Rol1994} by imposing a S/N greater than 6 for a given Gaussian fit to be usable. For each spectrum, the noise level was estimated from continuum fluctuations measured in empty channels. 

Figure 2 shows a high-quality spectrum obtained in the brightest parts of our mosaicked cube. This kind of spectrum is typical only for the regions of high emissivity of our FOV. Emission-line profiles corresponding to weaker areas are much noisier and complementary lines, found in the wavelength interval covered by the interference filter (see above), such as the $[$N\,\textsc{i}$]$\,$\lambda$6500 $\mbox{\AA}$ and $[$He\,\textsc{i}$]$\,$\lambda$6678 $\mbox{\AA}$ transitions are not discernible from noise fluctuations. The weakest line (i.e., $[$N\,\textsc{ii}$]$\,$\lambda$6548 $\mbox{\AA}$) in the example provided by Figure 2 has a S/N of 24.

\section{Results}

\subsection{Monochromatic maps}

Figures 3 to 7 provide the monochromatic, peak-intensity maps for all five emission lines investigated in this work. North is up and East is left in all figures. Figure 8 spectacularly displays the $\frac{[\textnormal{N}\,\textsc{ii}]\,\lambda6584}{\textnormal{H}{\alpha}\,\lambda6563}$ ratio throughout the whole FOV (see caption). Red emphasizes regions dominated by H{$\alpha$} while bluer shades pinpoint a stronger emission of the $[$N\,\textsc{ii}$]$\,$\lambda$6584 $\mbox{\AA}$ transition (with respect to the redder areas). Three stars with strong H{$\alpha$} lines are detected (the third being barely perceptible very close to the center of the figure). At the position of all other stars, bad pixels were removed and replaced by mean values statistically representing the surrounding nebular-gas content. This figure is particularly useful to describe ionized features in our FOV.

\begin{figure*}
  \begin{center}
    \leavevmode
    \scalebox{0.8}{\includegraphics[angle=90,scale=0.90]{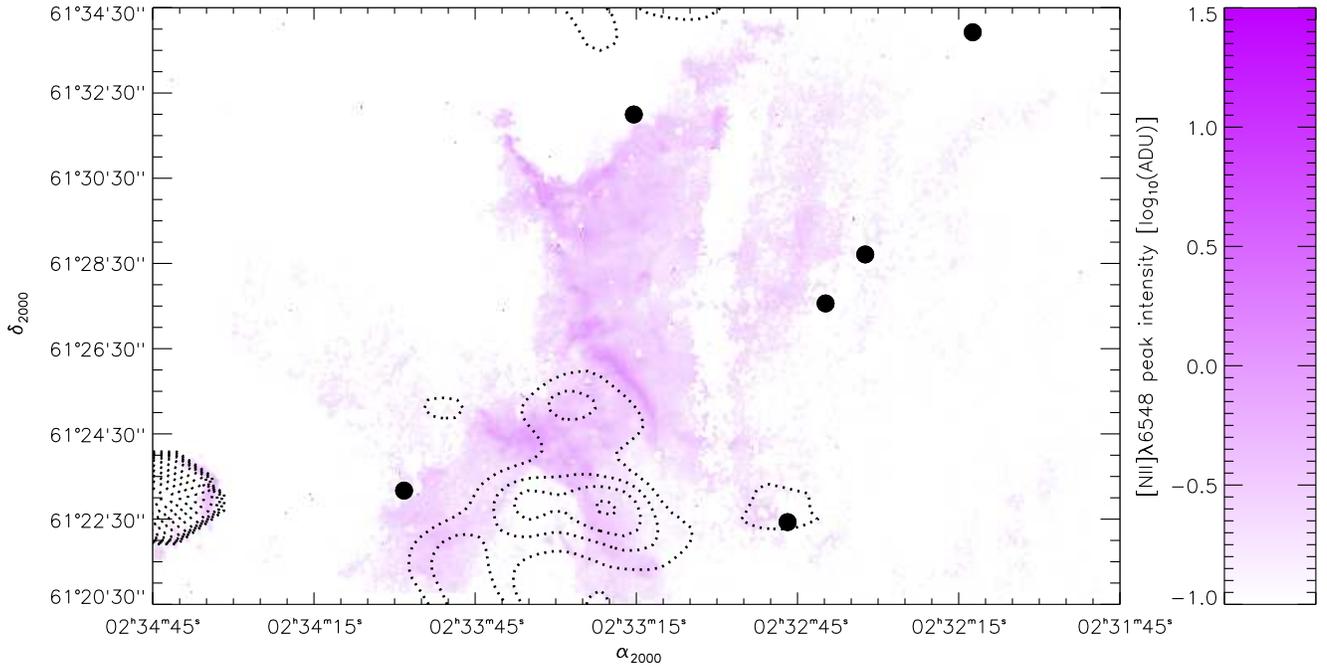}}
  \end{center}
  \caption{Peak-intensity map of the $[$N\,\textsc{ii}$]$\,$\lambda$6548 $\mbox{\AA}$ ionic transition in IC\,1805. Only emission-line profiles with S/N\,$>$\,6 were considered. A log$_{10}$-scale filter was applied in order to emphasize fainter ionized structures. As listed by \citet{Mas1995}, the filled circles indicate the position of the six O-stars found in the FOV. Dotted contours are traced out using an integrated intensity map of the molecular emission obtained from the FCRAO CO (1-0) observations. The spectral collapse of the molecular data cube was made for local standard of rest (LSR) radial velocities between -22.90 and -63.30 km s$^{-1}$, approximately corresponding to the extrema in velocities measured in the H$^{+}$ gas closest to Melotte 15 \citep{Lag2009a}. Eleven contour levels are drawn between 2 and 12 K km s$^{-1}$. The FCRAO CO (1-0) subcube has a noise level slightly below 1 K.}
\end{figure*}

\begin{figure*}
  \begin{center}
    \leavevmode
    \scalebox{0.8}{\includegraphics[angle=90,scale=0.90]{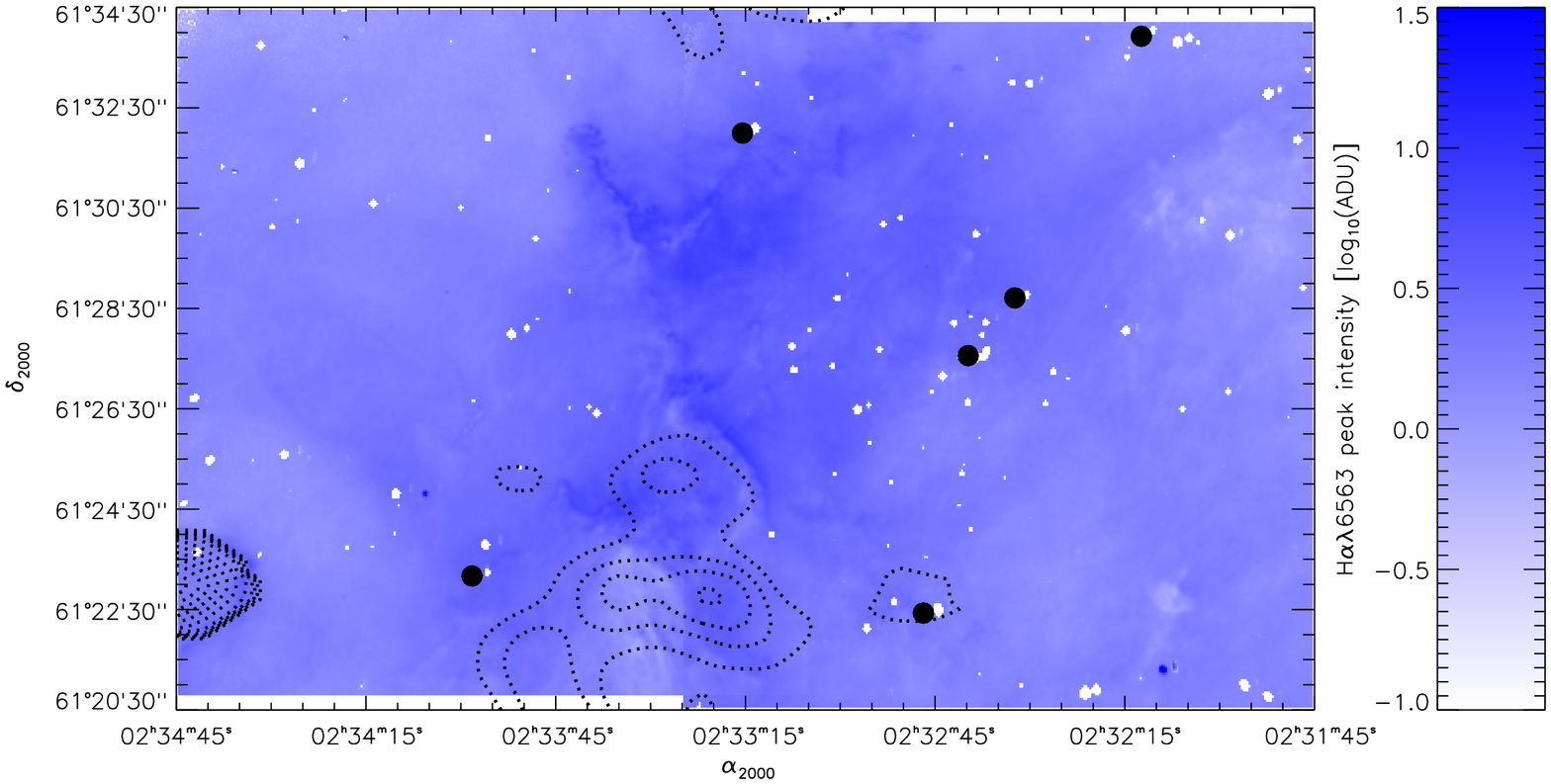}}
  \end{center}
  \caption{Peak-intensity map of the H{$\alpha$}\,$\lambda$6563 $\mbox{\AA}$ ionic transition in IC\,1805. More details are provided in the caption of Figure 3.}
\end{figure*}

\begin{figure*}
  \begin{center}
    \leavevmode
    \scalebox{0.8}{\includegraphics[angle=90,scale=0.90]{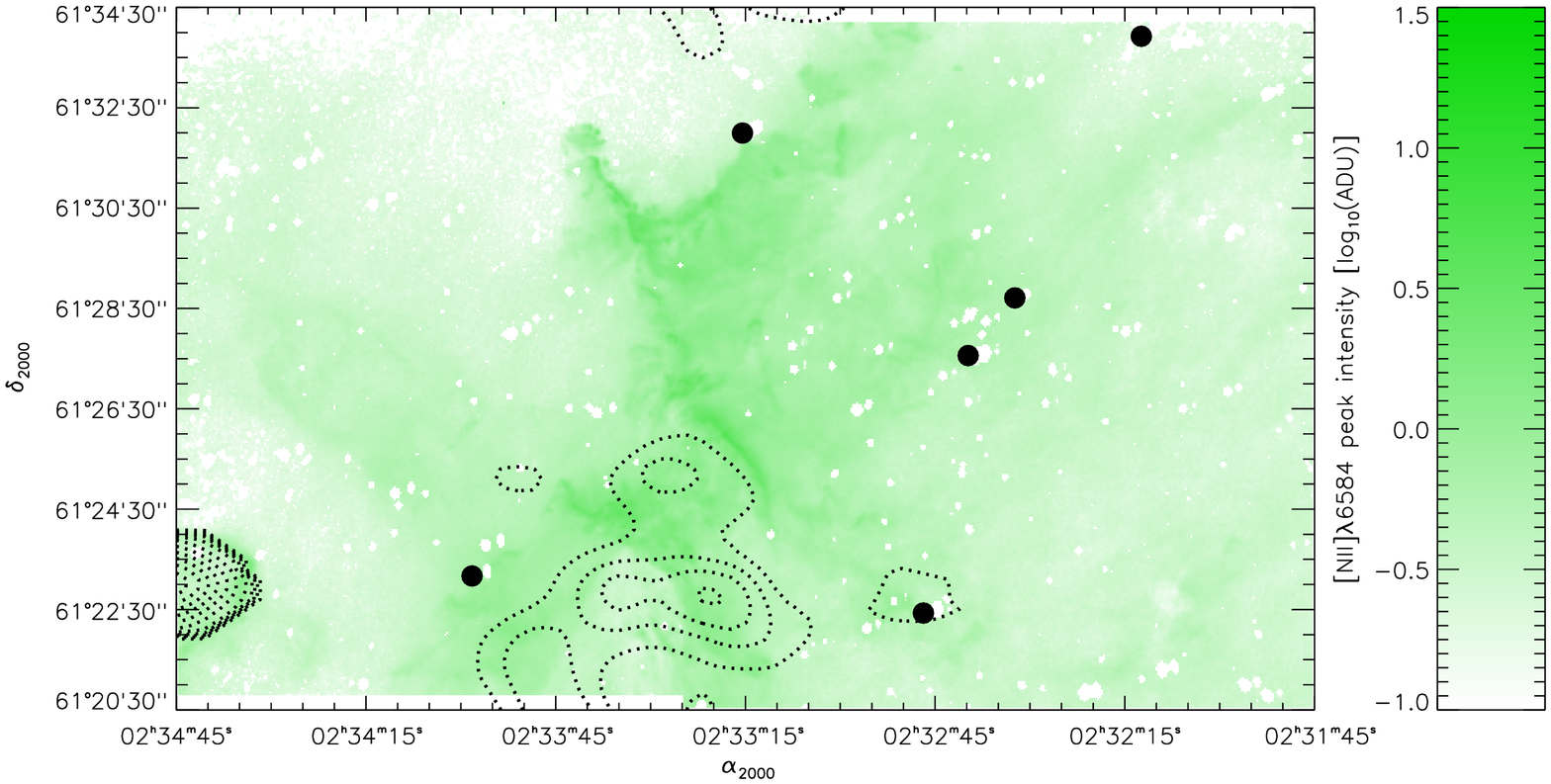}}
  \end{center}
  \caption{Peak-intensity map of the $[$N\,\textsc{ii}$]$\,$\lambda$6584 $\mbox{\AA}$ ionic transition in IC\,1805. More details are provided in the caption of Figure 3.}
\end{figure*}

\begin{figure*}
  \begin{center}
    \leavevmode
    \scalebox{0.8}{\includegraphics[angle=90,scale=0.90]{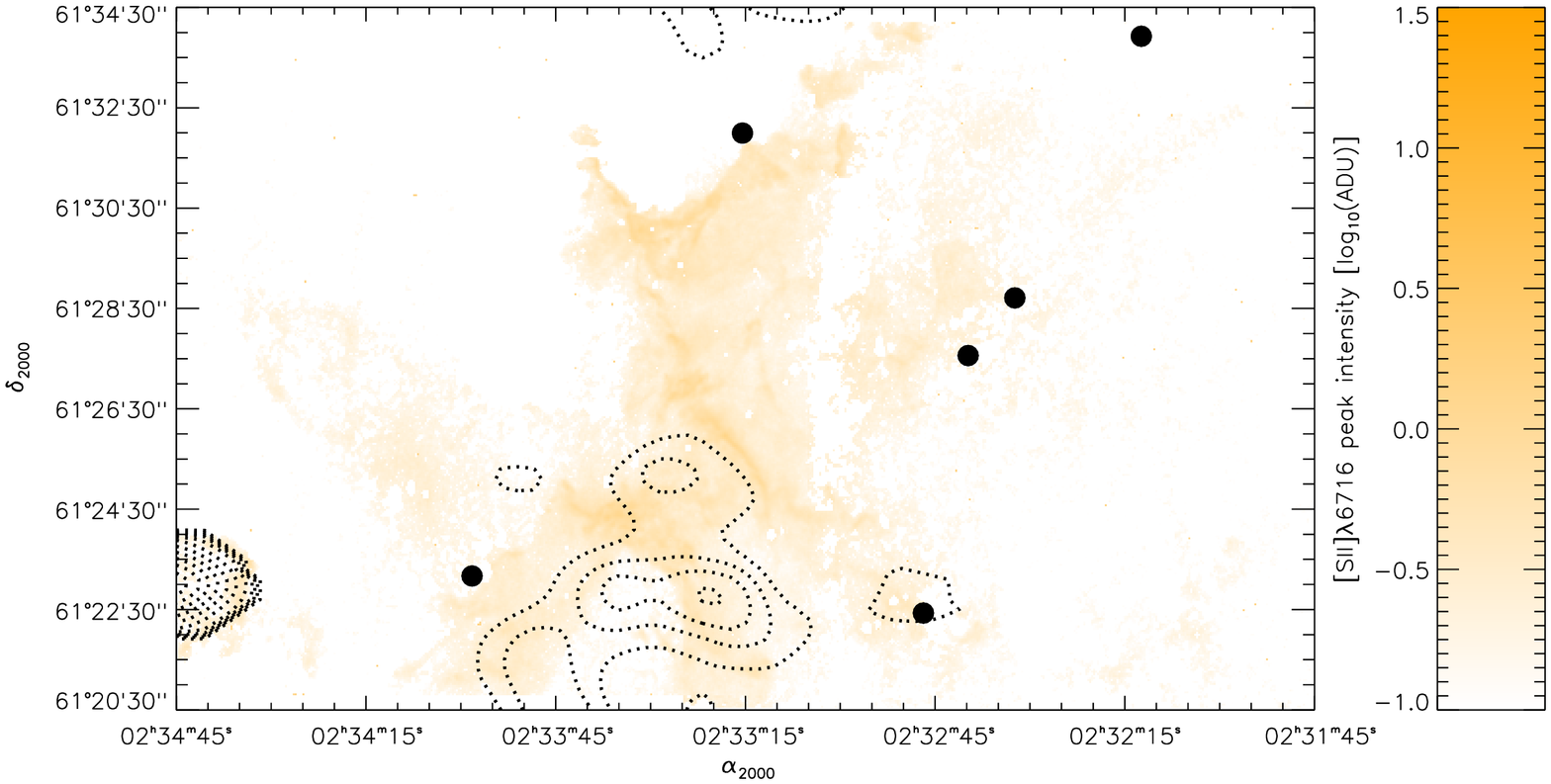}}
  \end{center}
  \caption{Peak-intensity map of the $[$S\,\textsc{ii}$]$\,$\lambda$6716 $\mbox{\AA}$ ionic transition in IC\,1805. More details are provided in the caption of Figure 3.}
\end{figure*}

\begin{figure*}
  \begin{center}
    \leavevmode
    \scalebox{0.8}{\includegraphics[angle=90,scale=0.90]{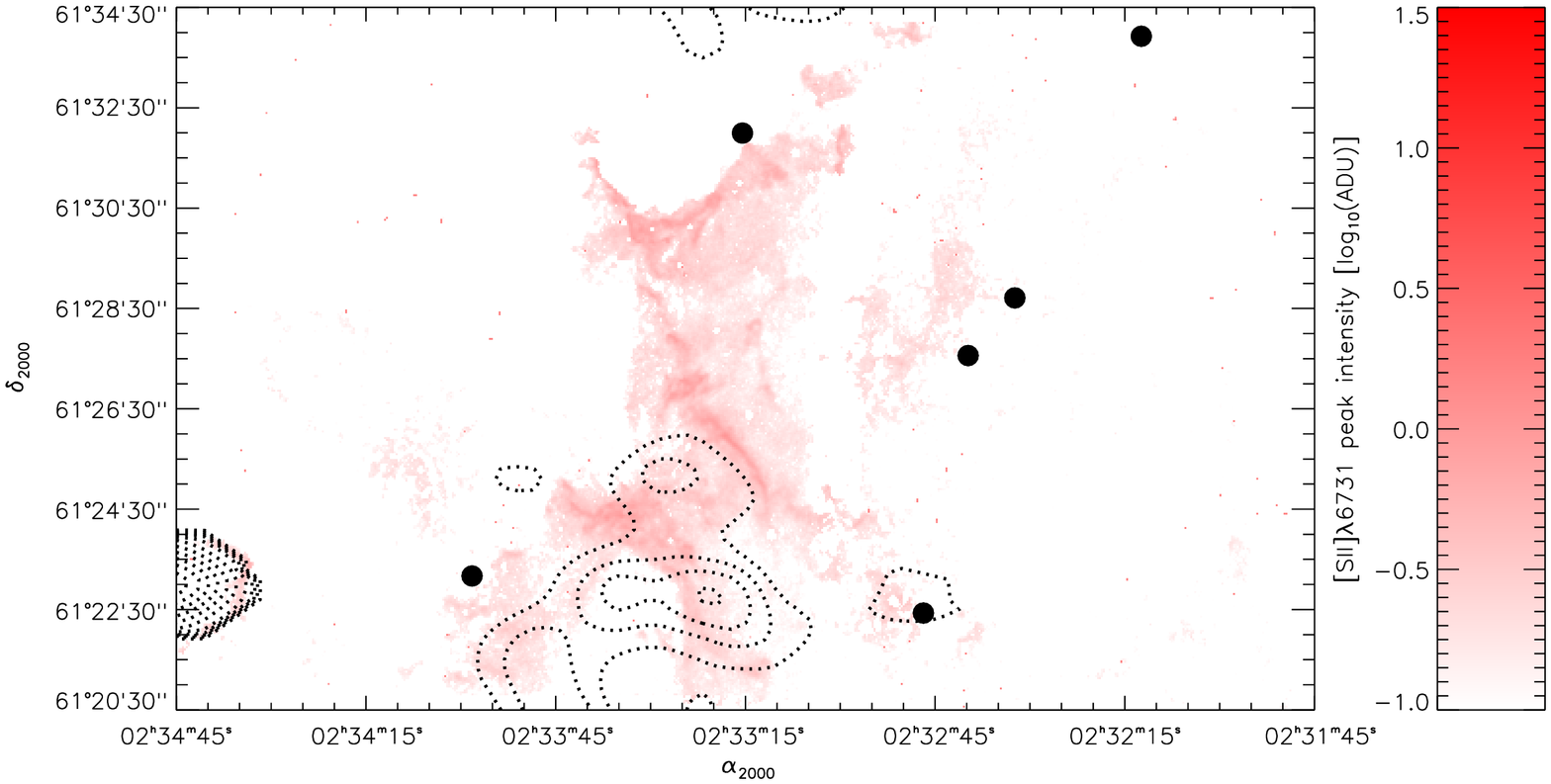}}
  \end{center}
  \caption{Peak-intensity map of the $[$S\,\textsc{ii}$]$\,$\lambda$6731 $\mbox{\AA}$ ionic transition in IC\,1805. More details are provided in the caption of Figure 3.}
\end{figure*}

\begin{figure*}
  \begin{center}
    \leavevmode
    \scalebox{0.8}{\includegraphics[scale=0.15]{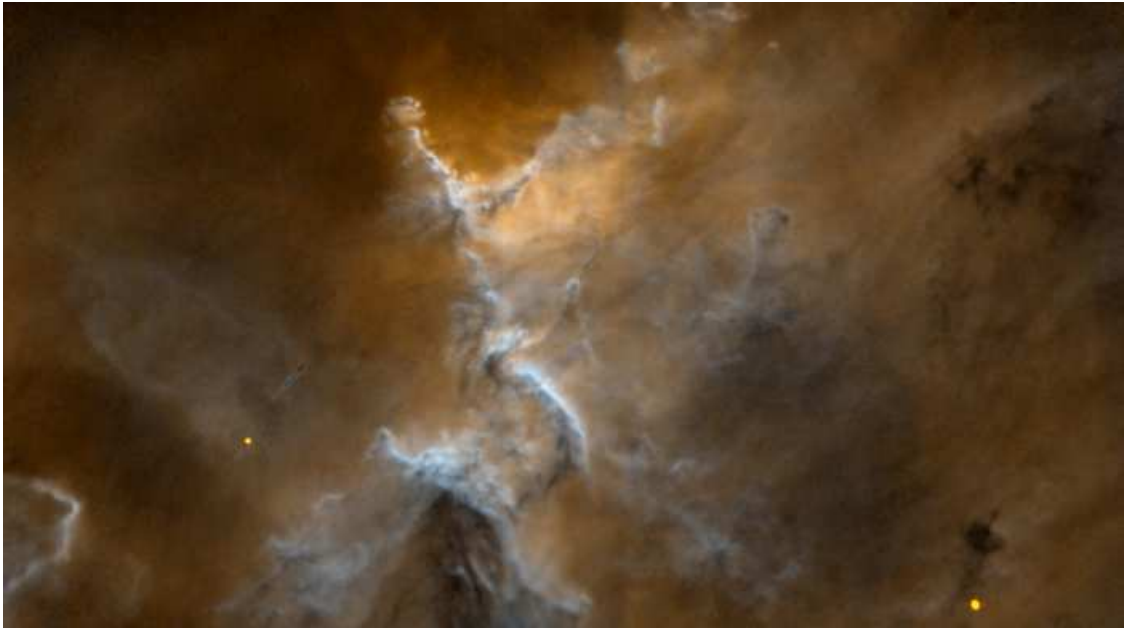}}
  \end{center}
  \caption{Image of the $\frac{[\textnormal{N}\,\textsc{ii}]\,\lambda6584}{\textnormal{H}{\alpha}\,\lambda6563}$ ratio in central IC\,1805. Red shades indicate that the H{$\alpha$} component clearly dominates the signal. Blue shades show the position of particularly bright $[$N\,\textsc{ii}$]$\,$\lambda$6584 $\mbox{\AA}$ emission. Stars were removed if not for three of them having well-defined, strong H{$\alpha$} emission (see text). This image was not processed using the information retrieved from the Gaussian fits but rather by simply collapsing the initial mosaicked cube in the channel intervals respectively enclosing the H{$\alpha$}\,$\lambda$6563 $\mbox{\AA}$ and $[$N\,\textsc{ii}$]$\,$\lambda$6584 $\mbox{\AA}$ ionic lines (e.g., see Figure 2).}
\end{figure*}

\begin{figure*}
  \begin{center}
    \leavevmode
    \scalebox{0.8}{\includegraphics[scale=0.45]{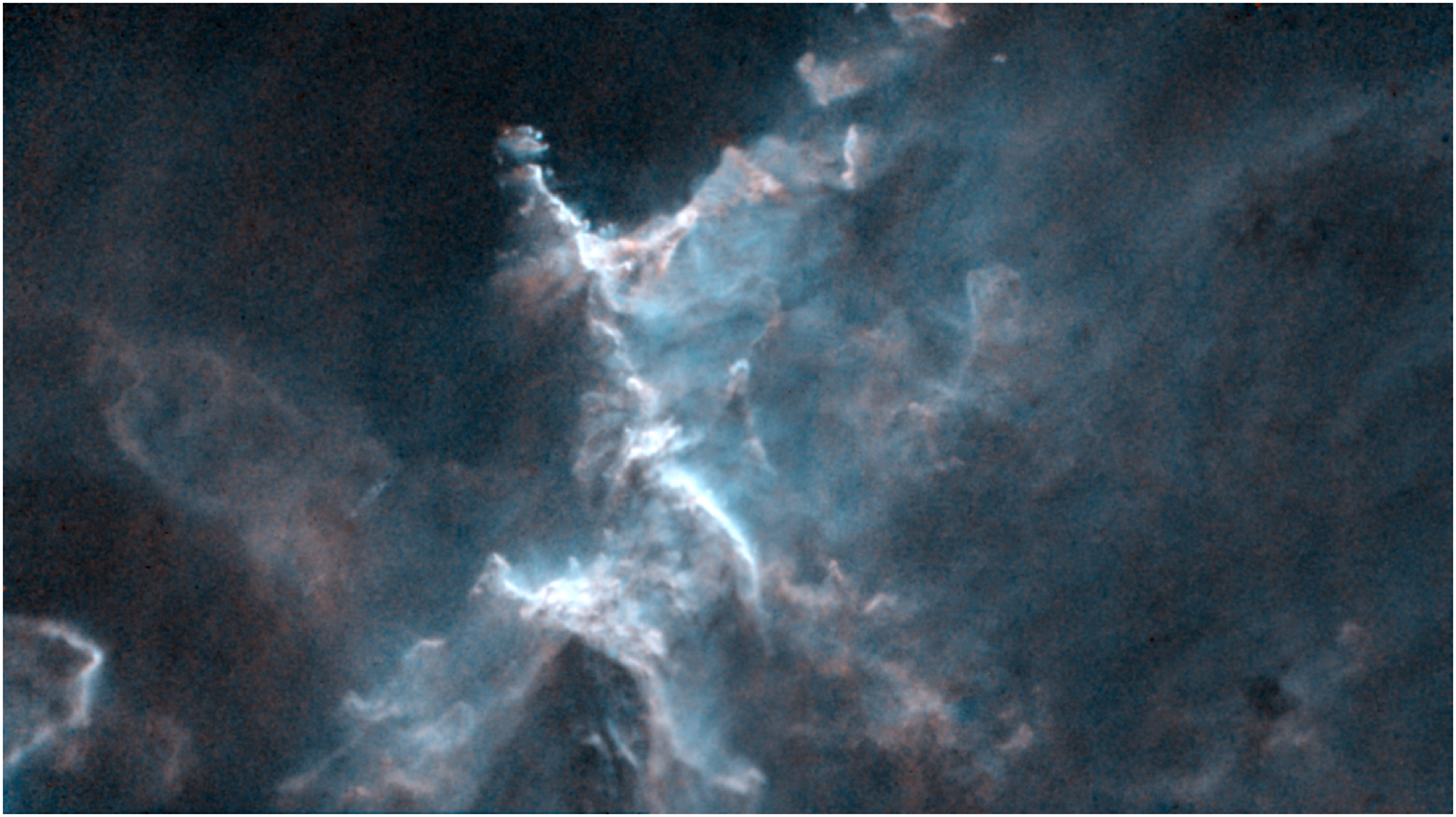}}
  \end{center}
  \caption{Image of the $\frac{[\textnormal{S}\,\textsc{ii}]\,\lambda\lambda6716, 6731}{[\textnormal{N}\,\textsc{ii}]\,\lambda6584}$ ratio in central IC\,1805. Blue shades indicate that the $[$N\,\textsc{ii}$]$ component clearly dominates the signal. Red shades show the position of particularly bright $[$S\,\textsc{ii}$]$ emission. The image was processed identically to Figure 8.}
\end{figure*}

The Five College Radio Astronomy Observatory (FCRAO) CO(1-0) survey of the 2$^{\mbox{\scriptsize{nd}}}$ Galactic quadrant \citep{Hey1998} has revealed very faint emission corresponding to tenuous molecular material at the position of the bright, central ionized structure in Figures 3 to 7. This suggests the presence of a large molecular fragment, surrounded by the most massive stars of the Melotte 15 cluster, that has undergone almost full erosion by the UV flux and stellar winds of the nearby ionizing sources. Figure 8 reveals the filamentary nature of the central structure. These filaments most likely trace out the spatial disposition, on the plane of the sky, of the eroded (sometimes fully ionized) molecular envelopes. The central structure is surrounded by diffuse ionized material likely associated to photoevaporated flows kinematically in agreement with the Champagne phase \citep{Lag2009a}. One of these flows (displaying blue shades in Figure 8) is particularly well-defined, propagating from the bottom-center of our FOV approximately toward the upper-left corner of Figure 8. West of the central structure, the spatial arrangement of the nebular gases is extremely complex. Ionized filaments are still perceptible although the contrast with the diffuse surroundings is not as clear. Toward the western boundary of Figure 8, even the H{$\alpha$} emission appears darker, obscured by interstellar dust. Shock excitation in the vicinity of the bright, central ionized structure, displayed in Figures 3 to 8, will be largely discussed and investigated in $\S$~5.2.1.

Toward the south-eastern portion of our FOV, a small CO fragment is clearly detected in the FCRAO survey at the systemic velocity of the IC\,1805 region (see CO contours in Figures 3 to 7). Its ionized counterpart is revealed by a thin, rounded ionization front especially visible in the lower-left corner of Figure 8 (see also Figure 19\textit{a}). This specific shape of the ionization front results from the vast majority of the ionizing sources in Melotte 15 being located behind the molecular clump. This was kinematically confirmed, in \citet{Lag2009a}, by the detection of an accelerated ionized outflow moving away from the observer (see $\S$~5.2.2.1). This flow eventually collides or simply coincides in lines-of-sight with the south-central/north-east flow mentioned in the previous paragraph. This results in particularly complicated kinematical motions in central IC\,1805 with H{$\alpha$} non-thermal line widths approaching the supersonic regime ($\sim$\,10 km s$^{-1}$) according to high-resolution observations \citep{Lag2009a}.

In the same area of Figure 8 (see also Figure 19\textit{c}), a cylindrical, cigar-like feature appears to be associated with an isolated star with strong H{$\alpha$} lines (see above). The feature has bright H$^{+}$ and N$^{+}$ rims but is almost completely gas-deprived near its center. This structure as well as the rounded ionization front found in the lower-left corner of Figure 8 will be discussed in $\S$~5.2.2. 

Figure 9 was processed identically to Figure 8 and displays the $\frac{[\textnormal{S}\,\textsc{ii}]\,\lambda\lambda6716, 6731}{[\textnormal{N}\,\textsc{ii}]\,\lambda6584}$ ratio (where the numerator is the sum of both lines of the $[$S\,\textsc{ii}$]$ doublet). Again, blue shades identify nitrogen-rich zones while red areas indicate that a sizeable role is played by sulfur in the overall gas emissivity. One immediately notices the complexity of the chemical properties in IC\,1805. While $[$N\,\textsc{ii}$]$ clearly dominates over $[$S\,\textsc{ii}$]$ where H{$\alpha$} is the brightest, weaker zones (in H{$\alpha$} integrated intensity) show an important increase of the relative contribution of the sulfur material. In particular, the reader's attention is directed toward the north-eastern filament of the bright, central structure described above. Series of small, quasi-circular blobs are found. The top three reveal two nitrogen-dominated features (in blue) and one with a stronger relative contribution in $[$S\,\textsc{ii}$]$ (in red). Investigating their corresponding spectrum, the $[$S\,\textsc{ii}$]$ intensity remains roughly constant from one blob to another while the $[$N\,\textsc{ii}$]$ lines suffer a strong attenuation along the line-of-sight of the ``redder'' one (which explains the relatively poor contribution of nitrogen towards it). This behavior is not generalized to our whole FOV although similar features can be found here and there.

The supernova remnant HB3 (see Figure 1) is located too far away from our FOV to have had a sizeable impact on the chemical properties in central IC\,1805. Alternatively, we can argue that the old, large molecular cloud, that gave birth to Melotte 15 roughly 2.5 Myr ago, may have had an inhomogeneous distribution of its chemical compounds. Old supernovae in IC\,1805, whose non-thermal emission has vanished since, could also be held responsible. We reiterate that the IC\,1805 region was most likely formed by a succession of different star clusters (see $\S$~2) i.e., although no indication for supernova remnants is currently found inside the large H\,\textsc{ii} region, it is highly probable that the inner zones of IC\,1805 were, at some time in the past, disturbed by supernova events associated to previous generations of massive stars. These stars could have had an intrinsic inhomogeneity in their inner nitrogen distribution which eventually led to an anisotropic dispersion of these chemical compounds, products of the CNO cycle, as each stellar object reached the end of its life (e.g., see the works by \citet{Mac2007}, \citet{Mac2008}, and \citet{Cha2010} on the Crab nebula).

\subsection{Diagnostic line ratios}

\subsubsection{Definition}

Line ratios, in the literature, usually correspond to ratios between two (or more) line fluxes. For a given emission line, the line flux is proportional to the product between its peak intensity and its width as returned by the Gaussian fit. In this work, the relatively low spectral resolution used (see $\S$~3) is roughly a factor 10 to 20 greater than typical non-thermal velocity fluctuations, along the line-of-sight, found in IC\,1805 using high-resolution observations of the H$^{+}$ kinematics \citep{Lag2009a}. Hence, in this work, the width of each emission line is entirely dominated by the instrumental response and the returned widths are very similar, from one ion to another, independently of the position in the FOV. Therefore, line ratios, in the following discussion, were strictly estimated using returned values for peak intensities while line widths were simply not considered.

\subsubsection{Line profile selection}

To assure the selection of the most reliable emission-line profiles in our sample of over 200\,000 spectra, a series of conditions are proposed. \textbf{All} of them need to be fulfilled otherwise the given spectrum is rejected from the discussion to follow. The conditions are summarized as follow:

\begin{enumerate}
\item The profile must show all five lines (i.e., H{$\alpha$}\,$\lambda$6563 $\mbox{\AA}$, $[$N\,\textsc{ii}$]$\,$\lambda$$\lambda$6548, 6584 $\mbox{\AA}$, and $[$S\,\textsc{ii}$]$\,$\lambda$$\lambda$6716, 6731 $\mbox{\AA}$) investigated in this work with reliable S/N above 6.
\newcounter{enumisaved1}
\setcounter{enumisaved1}{\value{enumi}}
\end{enumerate}

Two conditions are proposed in order to confirm that the kinematical properties of a given profile are physically acceptable\footnote{Although this paper is based on line-ratio measurements, kinematical properties were also carefully investigated. The question arises: could the line fluxes of kinematically ``unacceptable'' emission-line profiles (i.e., that do not fulfill Conditions (ii) and/or (iii)) be trusted? At this point, we have no reliable method to further investigate the question. We therefore chose to avoid any possible controversy/confusion on the matter and simply reject those kinematically peculiar spectra.}. More specifically, the $[$N\,\textsc{ii}$]$\,$\lambda$6548 $\mbox{\AA}$ and $[$N\,\textsc{ii}$]$\,$\lambda$6584 $\mbox{\AA}$ transitions must exhibit identical non-thermal motions since both lines are emitted by the same ions. The same argument also applies to the $[$S\,\textsc{ii}$]$\,$\lambda$6716 $\mbox{\AA}$ and $[$S\,\textsc{ii}$]$\,$\lambda$6731 $\mbox{\AA}$ lines.
 
\begin{enumerate}
\setcounter{enumi}{\value{enumisaved1}}
\item First, the difference, in centroid velocities, between both lines of a given doublet (i.e., $[$N\,\textsc{ii}$]$ or $[$S\,\textsc{ii}$]$) must not exceed 15 km s$^{-1}$. This corresponds roughly to 1/10th the spectral resolution of our observations (see $\S$~3) and therefore roughly accounts for the uncertainties on the Gaussian fits. Centroid-velocity differences, along a given line-of-sight, between the H$^{+}$, N$^{+}$, and S$^{+}$ material are allowed (and expected!) since all three ions are not necessarily co-spatial in central IC\,1805.
\item Secondly, the difference, in measured line widths, between all five components must not exceed 15 km s$^{-1}$. This is expected from all lines being dominated by the instrumental response (see $\S$~4.2.1). If observations with high spectral resolution (e.g., of the order of a few km s$^{-1}$) were used here, a condition similar to the previous one (on centroid velocities) would have been required (i.e., identical line widths, within the uncertainty bars of the Gaussian fits, between both lines of a given doublet, $[$N\,\textsc{ii}$]$ or $[$S\,\textsc{ii}$]$).
\newcounter{enumisaved2}
\setcounter{enumisaved2}{\value{enumi}}
\end{enumerate}

Finally, specific line ratios of the $[$N\,\textsc{ii}$]$ and $[$S\,\textsc{ii}$]$ transitions must agree with standard theoretical models of ionized nebulae.

\begin{enumerate}
\setcounter{enumi}{\value{enumisaved2}}
\item The computation of the $\frac{[\textnormal{S}\,\textsc{ii}]\,\lambda6716}{[\textnormal{S}\,\textsc{ii}]\,\lambda6731}$ line ratio must lead to a finite electron density. The upper and lower uncertainties on the density measurement (see $\S$~4.2.3) must also be finite.
\item The computation of the $\frac{[\textnormal{N}\,\textsc{ii}]\,\lambda6584}{[\textnormal{N}\,\textsc{ii}]\,\lambda6548}$ line ratio must not largely deviate from the theoretical value approaching 3 in low-density regimes typically found for H\,\textsc{ii} regions. Small deviations from 3 are expected (see $\S$~4.2.4) and explained (see $\S$~5.1) although values $\ll$\,2 or $\gg$\,4 would seem particularly hard to reconcile with the theory and would therefore demand to be rejected.
\end{enumerate}

All conditions taken in consideration, only 3\,057 emission-line profiles were retained, most of them being associated to the bright, central structure found in the overlapping region between the eastern and western field (see $\S$~5.2.1). Obviously, the summation of both cubes largely contributes to increase the data quality for duplicated pixels not only by increasing the peak signal for all lines but also by flattening noise fluctuations in empty channels (see Figure 2).

Considering the very large number of emission-line profiles available in our initial data set (more than 200\,000), we are fully aware that the conditions listed above drastically reduce the size of the retained sample. On the other hand, these conditions assure that the retained profiles are undoubtedly the most reliable available. 

We reiterate that a S/N greater than 6 is required to appropriately conduct an investigation on line ratios. However, a ``physical detection'' [of a given line] is considered when S/N\,$\gtrsim$\,3 \citep{Rol1994}. This will be used in a later subsection to investigate notable, although not necessarily reliable, line ratios for particular structures in IC\,1805 (see $\S$~5.2.2.2). 

\begin{figure*}
  \begin{center}
    \leavevmode
    \scalebox{0.8}{\includegraphics[scale=1.00]{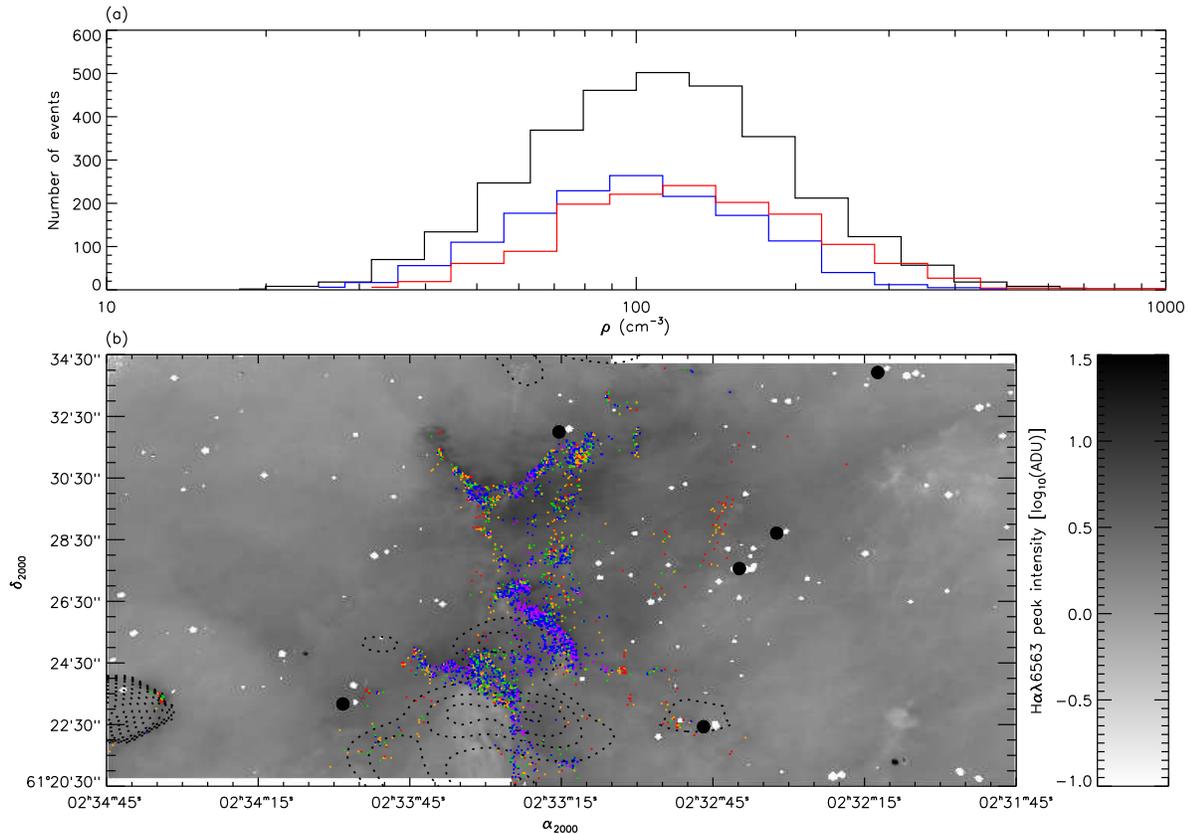}}
  \end{center}
  \caption{Panel (a): Histograms of the mean electron densities ($\rho$) in IC\,1805 computed from the $\frac{[\textnormal{S}\,\textsc{ii}]\,\lambda6716}{[\textnormal{S}\,\textsc{ii}]\,\lambda6731}$ line ratio. The black-line distribution corresponds to the 3\,057 emission-line profiles fulfilling all five conditions of $\S$~4.2.2. The blue curve is a subset of the black distribution, isolating the electron densities of specific points suggesting shock excitation. The red distribution uses the same points as the blue histogram although densities were computed from the post-subtraction $[$S\,\textsc{ii}$]$ doublets. The two colored histograms are discussed in $\S$~5.2.1 (see text). Panel (b): Spatial distribution of all 3\,057 electron-density measurements forming the black histogram of Panel (a). Densities between 1 and 75 cm$^{-3}$ (purple dots), 75 and 150 cm$^{-3}$ (blue dots), 150 and 200 cm$^{-3}$ (green dots), 200 and 300 cm$^{-3}$ (orange dots), and 300+ cm$^{-3}$ (red dots) are displayed.}
\end{figure*}

\subsubsection{Electron densities computed from the $[$S\,{\scriptsize{II}}$]$ doublet}

The $\frac{[\textnormal{S}\,\textsc{ii}]\,\lambda6716}{[\textnormal{S}\,\textsc{ii}]\,\lambda6731}$ line ratios were computed and used as input values for the Fivel \textsc{fortran} procedure \citep{DeR1987} adapted, for convenience, in \textsc{idl}. Assuming a constant electron temperature of 7\,400 K throughout IC\,1805 \citep{Lag2009a}, mean electron densities ($\rho$), in the S$^{+}$ ionic volume, were obtained for each nebular-gas column (i.e., line-of-sight).

Since uncertainties cannot be recovered from the use of the Fivel procedure, error bars on density measurements $\rho$ were computed as follows. For each line of the $[$S\,\textsc{ii}$]$ doublet, the uncertainty on the peak intensity, as returned by the Gaussian fit procedure, is provided by Equation 4\textit{a} of \citet{Lan1982}. From there, the statistical error on each line ratio is simply provided by the standard propagation of uncertainty. For each ratio, minimal and maximal plausible values are then easily obtained by respectively subtracting and adding this calculated uncertainty. Using these lower and upper limits on $\frac{[\textnormal{S}\,\textsc{ii}]\,\lambda6716}{[\textnormal{S}\,\textsc{ii}]\,\lambda6731}$, maximal ($\rho_{\textnormal{max}}$) and minimal ($\rho_{\textnormal{min}}$) densities, associated to $\rho$, can be obtained via Fivel. An asymmetrical error bar usually results where the upper (lower) uncertainty is provided by the difference between $\rho_{\textnormal{max}}$ ($\rho$) and $\rho$ ($\rho_{\textnormal{min}}$). Only $\rho$ values where both $\rho_{\textnormal{min}}$ and $\rho_{\textnormal{max}}$ are finite were retained (see $\S$~4.2.2). According to Fivel, this implies that $\rho_{\textnormal{min}}$ and $\rho_{\textnormal{max}}$ must be greater than 10 cm$^{-3}$ and lower than 15\,000 cm$^{-3}$ respectively (these two values mostly depend on our choice of a constant electron temperature of 7\,400 K in IC\,1805). Our method considers that the statistical error on $\rho$ is entirely dominated by the data quality (i.e., S/N of the $[$S\,\textsc{ii}$]$ lines) and that the uncertainties on the atomic parameters, used by Fivel, are negligible. However, using $\rho_{\textnormal{max}}$ and $\rho_{\textnormal{min}}$ to calculate the statistical error on $\rho$ probably overestimates the actual uncertainty on the electron density measurement.

\begin{figure*}
  \begin{center}
    \leavevmode
    \scalebox{0.8}{\includegraphics[scale=1.00]{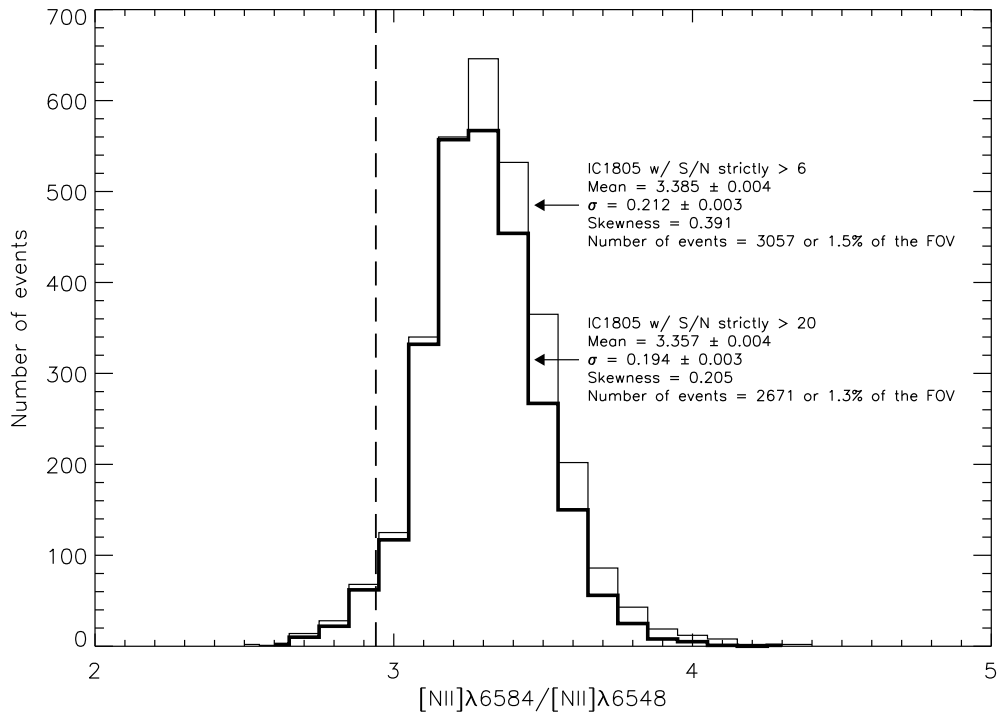}}
  \end{center}
  \caption{Histograms of the $\frac{[\textnormal{N}\,\textsc{ii}]\,\lambda6584}{[\textnormal{N}\,\textsc{ii}]\,\lambda6548}$ line ratios in IC\,1805. The theoretical value of 2.94, obtained in a low-density regime, is indicated by the long-dashed line. The regular-line distribution corresponds to the 3\,057 emission-line profiles fulfilling all five conditions of $\S$~4.2.2. The thick-line distribution considers only emission-line profiles with S/N greater than 20 (see text). Statistical properties of both histograms are provided within the figure itself.}
\end{figure*}

The distribution of the retained electron densities is provided by the black curve in Figure 10\textit{a} (the blue and red distributions will be discussed in $\S$~5.2.1). The histogram has a mean of 155$^{+\,135}_{-\,75}$ cm$^{-3}$ where the error bars correspond to the mean upper ($\left\langle \rho_{\textnormal{max}}\,-\,\rho \right\rangle$) and lower ($\left\langle \rho\,-\,\rho_{\textnormal{min}} \right\rangle$) uncertainty retrieved from the 3\,057 electron-density points preserved. Panel (b) provides the spatial distribution of these points, superimposed to a black-and-white reproduction of Figure 4. Purple dots have the lowest density values while red dots, the largest (see caption).

\begin{figure*}
  \begin{center}
    \leavevmode
    \scalebox{0.8}{\includegraphics[scale=1.00]{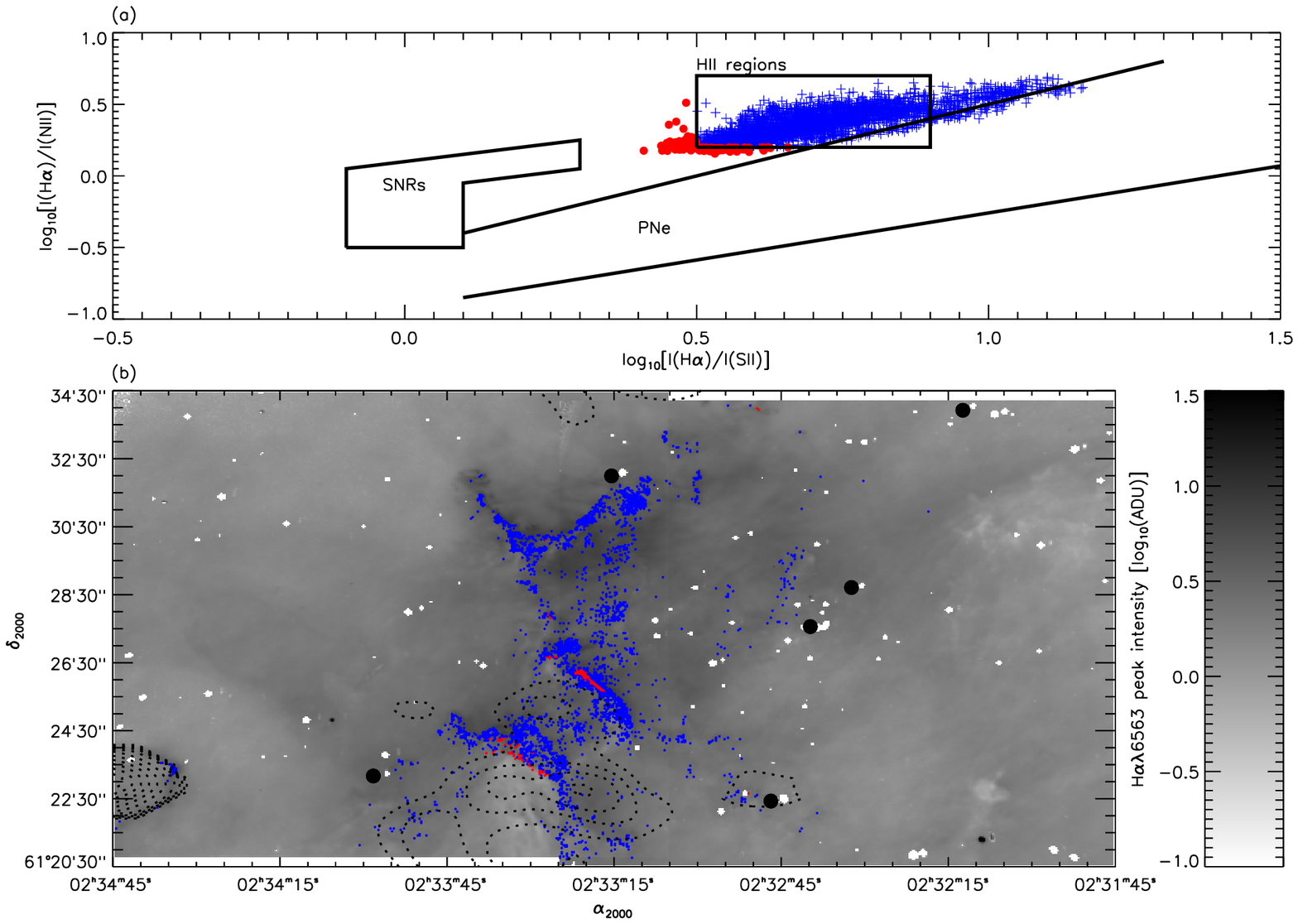}}
  \end{center}
  \caption{Panel (a): Diagnostic diagram of the log$_{10}\left[\frac{\textnormal{I}(\textnormal{H}\alpha)}{\textnormal{I}([\textnormal{S}\,\textsc{ii}])}\right]$ vs. log$_{10}\left[\frac{\textnormal{I}(\textnormal{H}\alpha)}{\textnormal{I}([\textnormal{N}\,\textsc{ii}])}\right]$ relation. By definition, $\textnormal{I}(\textnormal{H}\alpha)$ is the peak intensity of the H{$\alpha$}\,$\lambda$6563 $\mbox{\AA}$ line while $\textnormal{I}([\textnormal{S}\,\textsc{ii}])$ and \textnormal{I}([\textnormal{N}\,\textsc{ii}]) correspond respectively to the sum of both lines' peak intensity of the $[$S\,\textsc{ii}$]$ and $[$N\,\textsc{ii}$]$ doublet. While blue crosses indicate photoionization effects typically expected in H\,\textsc{ii} regions, red filled circles deviate toward the shock-dominated regime of SNRs (see text). Panel (b): Spatial distribution of all 3\,057 points found in Panel (a). Blue crosses are represented by blue dots and red filled circles, by red dots.}
\end{figure*}

\begin{figure*}
  \begin{center}
    \leavevmode
    \scalebox{0.8}{\includegraphics[scale=1.00]{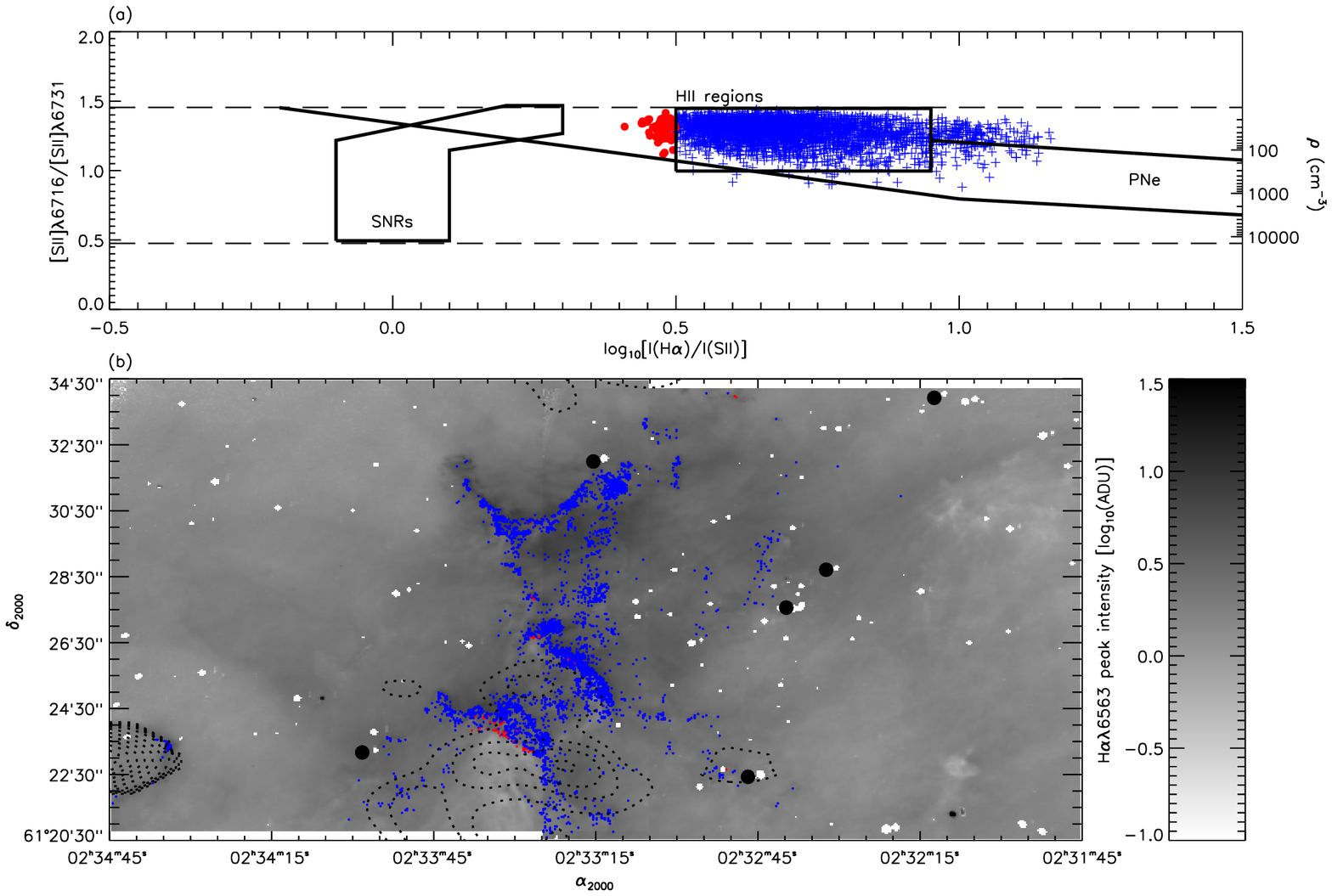}}
  \end{center}
  \caption{Panel (a): Diagnostic diagram of the log$_{10}\left[\frac{\textnormal{I}(\textnormal{H}\alpha)}{\textnormal{I}([\textnormal{S}\,\textsc{ii}])}\right]$ vs. $\frac{[\textnormal{S}\,\textsc{ii}]\,\lambda6716}{[\textnormal{S}\,\textsc{ii}]\,\lambda6731}$ line ratio relation. The right-hand ordinate indicates the correspondence between $[$S\,\textsc{ii}$]$ line ratios and electron densities $\rho$ as provided by our \textsc{idl} adaptation of the Fivel \textsc{fortran} procedure. Panel (b): Spatial distribution of all preserved points. More details are provided in the caption of Figure 12.}
\end{figure*}

\begin{figure*}
  \begin{center}
    \leavevmode
    \scalebox{0.8}{\includegraphics[scale=1.00]{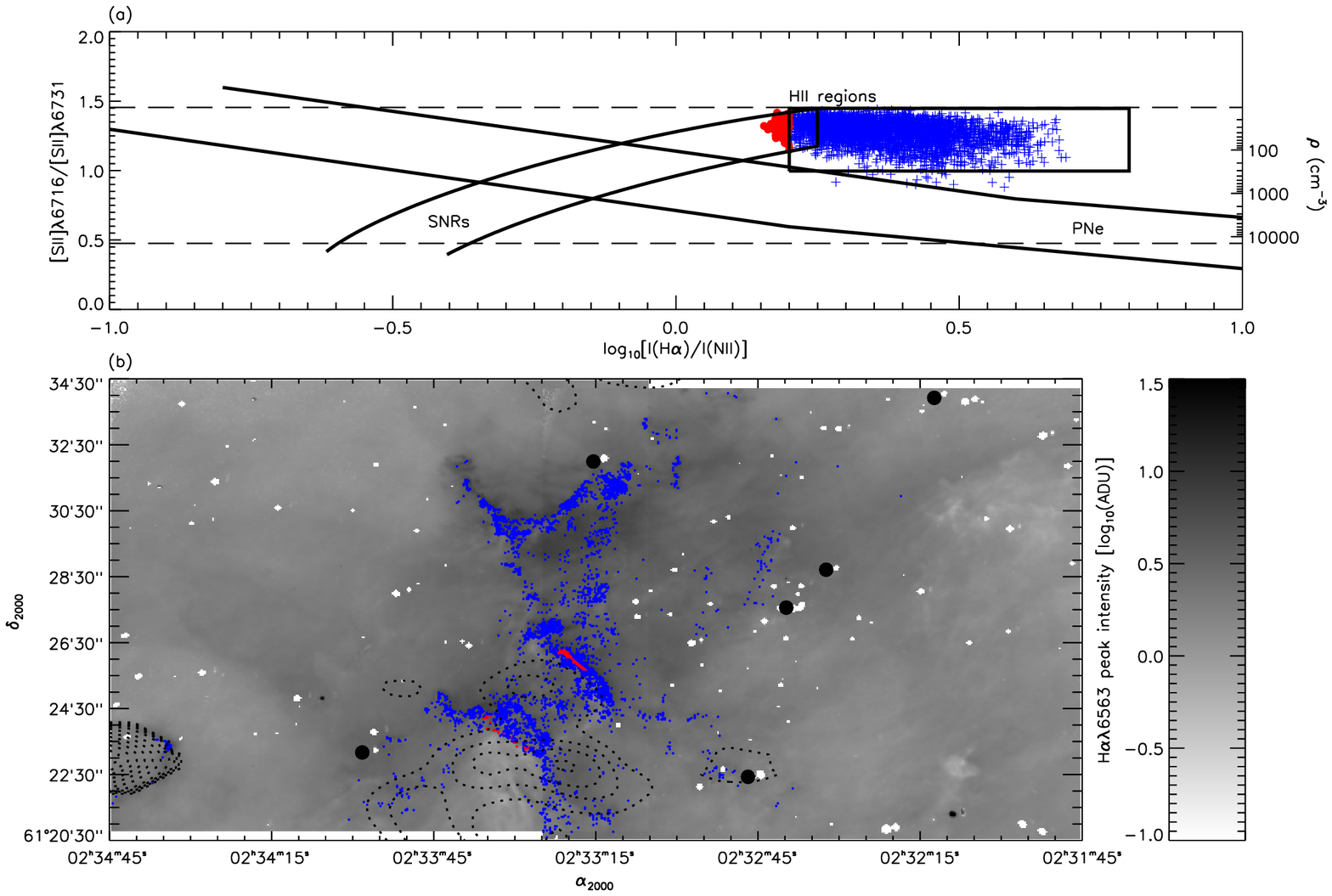}}
  \end{center}
  \caption{Diagnostic diagram of the log$_{10}\left[\frac{\textnormal{I}(\textnormal{H}\alpha)}{\textnormal{I}([\textnormal{N}\,\textsc{ii}])}\right]$ vs. $\frac{[\textnormal{S}\,\textsc{ii}]\,\lambda6716}{[\textnormal{S}\,\textsc{ii}]\,\lambda6731}$ line ratio relation. Panel (b): Spatial distribution of all preserved points. More details are provided in the caption of Figures 12 and 13.}
\end{figure*}

\begin{figure*}
  \begin{center}
    \leavevmode
    \scalebox{0.8}{\includegraphics[scale=1.10]{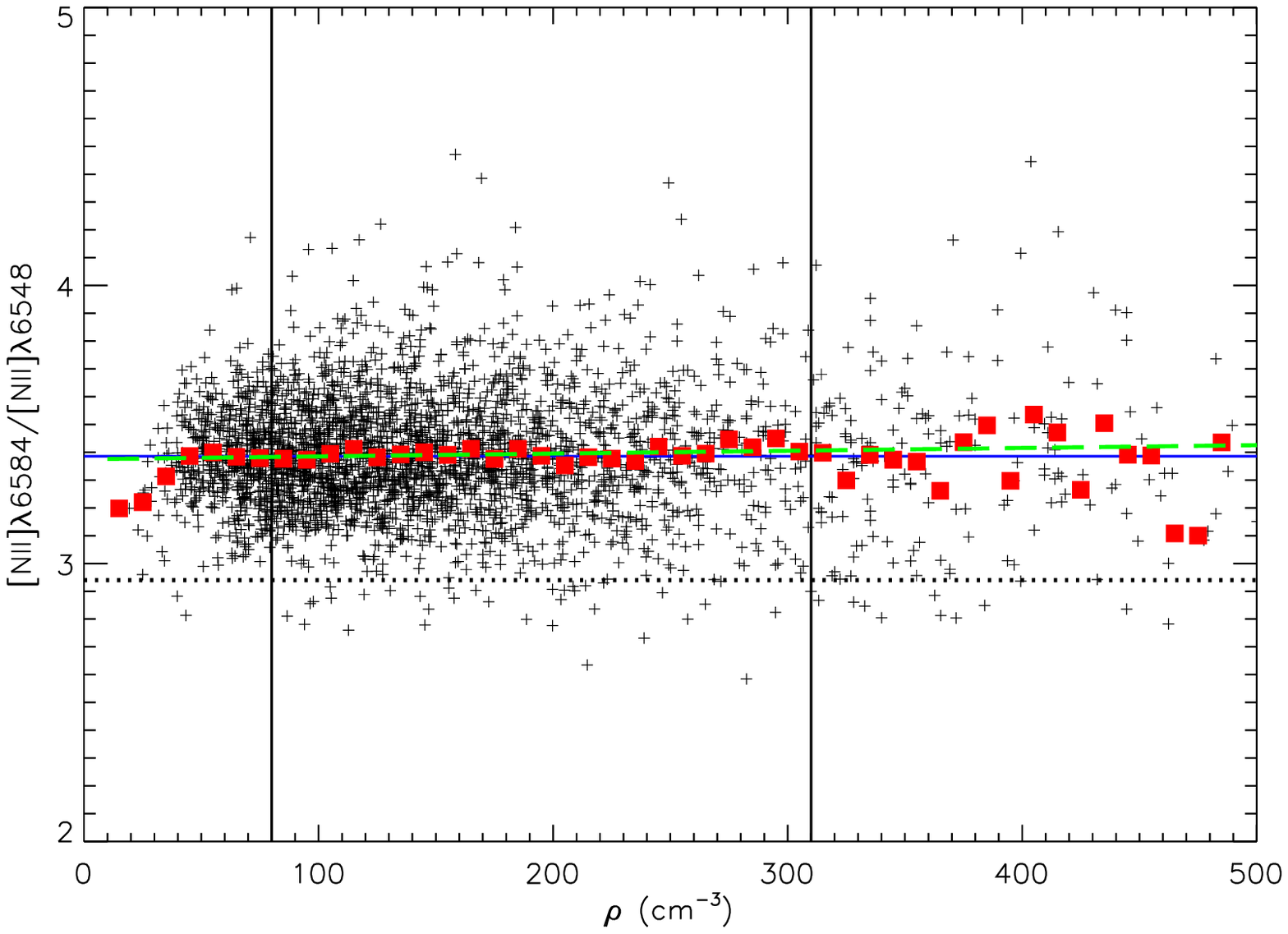}}
  \end{center}
  \caption{Diagram of the mean electron density ($\rho$) vs. $\frac{[\textnormal{N}\,\textsc{ii}]\,\lambda6584}{[\textnormal{N}\,\textsc{ii}]\,\lambda6548}$ line ratio relation. The horizontal dotted line indicates the theoretical value of the $\frac{[\textnormal{N}\,\textsc{ii}]\,\lambda6584}{[\textnormal{N}\,\textsc{ii}]\,\lambda6548}$ ratio as predicted in a low-density regime. The horizontal regular line marks the mean $\frac{[\textnormal{N}\,\textsc{ii}]\,\lambda6584}{[\textnormal{N}\,\textsc{ii}]\,\lambda6548}$ ratio retrieved from Figure 11 (regular-line distribution). The vertical regular lines isolate the 80$-$310 cm$^{-3}$ interval in densities where collisional de-excitations are expected to impact the $[$N\,\textsc{ii}$]$\,$\lambda$6548 $\mbox{\AA}$ line while the $[$N\,\textsc{ii}$]$\,$\lambda$6584 $\mbox{\AA}$ line remains theoretically unaffected (see text). Red filled squares correspond to the mean $\frac{[\textnormal{N}\,\textsc{ii}]\,\lambda6584}{[\textnormal{N}\,\textsc{ii}]\,\lambda6548}$ ratio observed for each bin of 10 cm$^{-3}$ in densities. The dashed green line indicates the linear fit to the sample of filled squares found between 80 and 310 cm$^{-3}$. Large fluctuations of the filled-square symbols at high densities ($\rho$\,$\gtrsim$\,350 cm$^{-3}$) result from low amounts of points (black crosses) within each bin.}
\end{figure*} 

\subsubsection{The $[$N\,{\scriptsize{II}}$]$ line ratio}

Figure 11 provides the distribution of the $\frac{[\textnormal{N}\,\textsc{ii}]\,\lambda6584}{[\textnormal{N}\,\textsc{ii}]\,\lambda6548}$ line ratio in the vicinity of the Melotte 15 star cluster. The vertical long-dashed line indicates the theoretical value as suggested by the ratio of the transition probabilities of both ionic lines in a low-density regime \citep[Chapter 5]{Ost2006}. All 3\,057 points mentioned in $\S$~4.2.3 formed the regular-line distribution. As shown by the thick-line histogram, the presence of large values i.e., above $\sim$3.5$-$3.6, is reduced by considering only data points of very high quality (i.e., requiring here, arbitrarily, that S/N\,$>$\,20 in the first condition of $\S$~4.2.2). Uncertainties on the mean and one-standard deviation of both histograms are statistical uncertainties retrieved assuming normal distributions \citep[Chapter 5]{Bev1969}. Extreme values for $\frac{[\textnormal{N}\,\textsc{ii}]\,\lambda6584}{[\textnormal{N}\,\textsc{ii}]\,\lambda6548}$ i.e., below 2 or well-above 4, were not measured. Hence, Condition 5, listed in $\S$~4.2.2, was not used in this work to reject particular emission-line profiles although it could used by other authors for future, similar studies. 

Considering the relatively high S/N used here (above 20 for the thick-line histogram of Figure 11 hence suggesting a very good signal quality), deviations from $\frac{[\textnormal{N}\,\textsc{ii}]\,\lambda6584}{[\textnormal{N}\,\textsc{ii}]\,\lambda6548}$\,$\sim$\,3 seem genuine and will be addressed in $\S$~5.1.

\subsubsection{Line-ratio diagnostic diagrams}

Panels (a) of Figures 12 to 14 provide a series of line-ratio diagnostic diagrams commonly used in optical observations. Each diagram contains the 3\,057 points already discussed in $\S$$\S$~4.2.3 and 4.2.4. Areas labeled ``H\,\textsc{ii} regions'', ``SNRs'', and ``PNe'' were all reproduced according to Figures 1 to 3 of \citet{Sab1977}. Long-dashed lines, in Figures 13\textit{a} and 14\textit{a} indicate the lower and upper limits for $\frac{[\textnormal{S}\,\textsc{ii}]\,\lambda6716}{[\textnormal{S}\,\textsc{ii}]\,\lambda6731}$ that provide finite values of electron densities $\rho$ given a constant temperature of 7\,400 K in IC\,1805 (see $\S$~4.2.3).

In all three diagrams, crosses, colored in blue, are used to represent points found (in majority) within the ``H\,\textsc{ii} regions'' area. The relatively high numbers of crosses is not surprising for IC\,1805, already cataloged as an H\,\textsc{ii} region. A certain fraction of the sample, however, falls outside this area. These points are symbolized as red filled circles and point approximately toward the ``SNRs'' area, hence suggesting evidence for shock excitation in the targeted ISM volume. This will be discussed in $\S$~5.2. Out of the 3\,057 points retained for this study, respectively 134 (4.4\%), 67 (2.2\%), and 84 (2.7\%) are displayed as red filled circles in Figures 12\textit{a} to 14\textit{a}.

Panels (b), in all three figures, provide the spatial distribution of all crosses (shown as blue dots) and circles (shown as red dots) displayed in the diagnostic diagram of their respective Panel (a). As in Figure 10\textit{b}, a black-and-white reproduction of Figure 4 was used. Red dots in Figures 12\textit{b} to 14\textit{b} all point at the same areas and reveal that the southern portions of the bright, central structure ($\delta_{2000}$\,$<$\,61$\arcdeg$26$\arcmin$30$\arcsec$) may encompassed material subject to ionization by shocks.
 
\section{Discussion}

\subsection{The $[$N\,{\scriptsize{II}}$]$ line ratio: deviations from the theoretical value}

As demonstrated in $\S$~4.2.4, values for the $\frac{[\textnormal{N}\,\textsc{ii}]\,\lambda6584}{[\textnormal{N}\,\textsc{ii}]\,\lambda6548}$ line ratio above 3.5 and approaching 4 are partially attributed to poorer data quality especially affecting the lower-signal, noisy $[$N\,\textsc{ii}$]$\,$\lambda$6548 $\mbox{\AA}$ line (e.g., see Figure 2). Figure 11 indicates that the use of profiles whose emission lines are strictly characterized by ``very'' high S/N (in our case, greater than 20) contributes to attenuate the right tail of the distribution. The asymmetry coefficient (i.e., skewness) is reduced from roughly 0.4 to 0.2. Consequently, the width of the distribution is statistically narrower while the mean of the histogram slightly varied toward the expected value of $\sim$3 for $\frac{[\textnormal{N}\,\textsc{ii}]\,\lambda6584}{[\textnormal{N}\,\textsc{ii}]\,\lambda6548}$.

Nonetheless, independently of the threshold in S/N used (e.g., 6 or 20), Figure 11 reveals that the mean $\frac{[\textnormal{N}\,\textsc{ii}]\,\lambda6584}{[\textnormal{N}\,\textsc{ii}]\,\lambda6548}$ ratio exceeds, by a few tenths, the theoretical value. We propose that this could be attributed to a simple statistical effect by which the ratio of transition probabilities between the $[$N\,\textsc{ii}$]$\,$\lambda$6548 $\mbox{\AA}$ and $[$N\,\textsc{ii}$]$\,$\lambda$6584 $\mbox{\AA}$ transitions does not necessarily equal 3 given the physical properties governing the IC\,1805 nebula. Note that the curve of transmission of the interference filter used for data acquisition has not revealed any significant difference between the transmission coefficients of both lines of the $[$N\,\textsc{ii}$]$ doublet.

Table 3.15 of \citet{Ost2006} provides, for different ionic transitions, the critical electron density above which collisional de-excitations, caused by free electrons, become important. This density is estimated at 80 and 310 cm$^{-3}$ respectively for the $[$N\,\textsc{ii}$]$\,$\lambda$6548 $\mbox{\AA}$ and $[$N\,\textsc{ii}$]$\,$\lambda$6584 $\mbox{\AA}$ transitions. Hence, for a given electron density between these two values and characterizing the N$^{+}$ volume, the $[$N\,\textsc{ii}$]$\,$\lambda$6548 $\mbox{\AA}$ line intensity will be less than expected, a fraction of its flux being ``lost'' in (non-radiative) collisional de-excitations. On the other hand, the $[$N\,\textsc{ii}$]$\,$\lambda$6584 $\mbox{\AA}$ line intensity will not be affected until densities reach the 310 cm$^{-3}$ plateau. Figure 10\textit{a} provides reasonable evidences for electron densities between 80 and 310 cm$^{-3}$ in IC\,1805. Hence, a mean $\frac{[\textnormal{N}\,\textsc{ii}]\,\lambda6584}{[\textnormal{N}\,\textsc{ii}]\,\lambda6548}$ ratio of 3.3$-$3.4 could be explained by a $[$N\,\textsc{ii}$]$\,$\lambda$6548 $\mbox{\AA}$ line partially experiencing collisional de-excitations. 

From the results presented in $\S$$\S$~4.2.3 and 4.2.4 (for S/N\,$>$\,6), Figure 15 provides the electron densities (as measured from the $\frac{[\textnormal{S}\,\textsc{ii}]\,\lambda6716}{[\textnormal{S}\,\textsc{ii}]\,\lambda6731}$ line ratio; see Figure 10) vs. the $\frac{[\textnormal{N}\,\textsc{ii}]\,\lambda6584}{[\textnormal{N}\,\textsc{ii}]\,\lambda6548}$ line ratio (see Figure 11) relation. Each $\rho$\,$-$\,$\frac{[\textnormal{N}\,\textsc{ii}]\,\lambda6584}{[\textnormal{N}\,\textsc{ii}]\,\lambda6548}$ pair is represented by a black cross. For each bin of 10 cm$^{-3}$ between 10 and 500 cm$^{-3}$, the mean $\frac{[\textnormal{N}\,\textsc{ii}]\,\lambda6584}{[\textnormal{N}\,\textsc{ii}]\,\lambda6548}$ line ratio is displayed as a red filled square. A standard linear regression \citep[Chapter 4]{Bev1969} was applied to the filled-square sample, in the 80$-$310 cm$^{-3}$ interval, and is displayed as the dashed green line. The value of each bin was adequately weighted in the regression i.e., bins with large numbers of crosses were given a more important statistical weight. 

A clear, monotonic increase of the $\frac{[\textnormal{N}\,\textsc{ii}]\,\lambda6584}{[\textnormal{N}\,\textsc{ii}]\,\lambda6548}$ line ratio with increasing densities would be expected as the $[$N\,\textsc{ii}$]$\,$\lambda$6548 $\mbox{\AA}$ line becomes more-and-more affected by collisional de-excitations (i.e., decreases in intensity). As expected, the linear relation shows a positive slope although its value, very close to 0, largely suggests no correlation at all. Indeed, the quality of the linear fit is relatively poor with a correlation coefficient of 0.35. Hence, we believe that no conclusion can be drawn from Figure 15. Two possibilities are however provided in order to explain such behavior. 

First, the reader should note that the uncertainties are relatively large for both parameters investigated. The mean uncertainty for the $\frac{[\textnormal{N}\,\textsc{ii}]\,\lambda6584}{[\textnormal{N}\,\textsc{ii}]\,\lambda6548}$ line ratio is 0.17 while typical uncertainties for the electron-density measurements $\rho$ are provided in the last paragraph of $\S$~4.2.3. The poor correlation of the linear fit in Figure 15 could be attributed to large error bars. 

Moreover, the absence of a well-defined correlation between $\rho$ and $\frac{[\textnormal{N}\,\textsc{ii}]\,\lambda6584}{[\textnormal{N}\,\textsc{ii}]\,\lambda6548}$ would not be surprising if the N$^{+}$ and S$^{+}$ material are not perfectly cospatial in the vicinity of Melotte 15. Given a difference of more than 4 eV between the ionization potentials of neutral nitrogen and sulfur, it can be assumed that the electron densities, computed from the $[$S\,\textsc{ii}$]$ doublet, may not reflect accurately the physical conditions prevailing in the N$^{+}$ ionic volume. Therefore, the $\rho$ vs. $\frac{[\textnormal{N}\,\textsc{ii}]\,\lambda6584}{[\textnormal{N}\,\textsc{ii}]\,\lambda6548}$ diagram of Figure 15 could be an interesting diagnostic tool although our data set may not provide the appropriate spectral information to reliably construct such a diagram. Alternatively, given the first ionization potential of nitrogen (14.5 eV) and oxygen (13.6 eV), the $\frac{[\textnormal{O}\,\textsc{ii}]\,\lambda3726}{[\textnormal{O}\,\textsc{ii}]\,\lambda3729}$ line ratio would likely give a more accurate indication of the density behavior inside the N$^{+}$ volume.

\subsection{Shock excitation in IC\,1805}

Panels (a) of Figures 12 to 14 reveal that a large fraction of the nebular-gas content in the vicinity of the Melotte 15 star cluster appears to be dominated by photoionization, typical of standard H\,\textsc{ii} regions. However, points identified outside the expected regime for H\,\textsc{ii} regions and located near the ``SNRs'' area demand to be closely investigated (see $\S$~4.2.5). In all three figures, these are symbolized as red filled circles and potentially indicate the presence of shocks in the targeted nebular volume. These areas of our FOV, suggesting shock excitation, are identified as red dots in all three Panels (b) of the same figures.

However, at this point, shock excitation in IC\,1805 remains highly hypothetical. This is somewhat surprising considering the presence of relatively massive stars in Melotte 15 and their associated strong stellar winds (see $\S$~2). In order to deeply investigate the impact of shock excitation in IC\,1805, we assume that, if present, the shock-excited ionized material is relatively confined and localized along the line-of-sight i.e., considering the imposing dimensions of the large H\,\textsc{ii} region (e.g., \citealt{Lag2009a,Lag2009b}), the interface between compressive shocks and the surrounding ISM could be several times smaller than the size of the nebula itself. This said, the expected spectral signature of shock excitation could be strongly diluted by photoionized foreground/background material.

\subsubsection{Central structure}

A total of 2\,378 out of 3\,057 emission-line profiles retained for this study are directly associated to the bright, central structure clearly visible in Figure 8. The strong emission of its ionized component (see peak intensities in Figures 3 to 7) and therefore the quality of the gathered signal make this structure, surrounded by the most massive stars of the Melotte 15 cluster, the ideal feature in our quest for shock excitation in central IC\,1805. 

First, a series of four weaker portions [in gas emission], specifically surrounding the central structure, were selected. Each of these portions was spatially binned into a single 1\,$\times$\,1\,$\times$\,249 (spatial\,$\times$\,spatial\,$\times$\,channels) emission-line profile\footnote{The spatial binning of numerous emission-line profiles into a single one flattens out noise fluctuations. Therefore, the resulting spectrum is of very high quality.}. For each of the 2\,378 spectra selected and investigated in this subsection, the corresponding foreground/background emission was approximately recovered using a linear combination of these four 1\,$\times$\,1\,$\times$\,249 profiles. Each of the four linear coefficients was weighted via the inverse of the distance separating the targeted pixel from the center of the corresponding weaker portion (e.g., the larger this distance is, the smaller is the corresponding linear coefficient and, therefore, the smaller is the statistical impact of this weaker zone on the computation of the foreground/background spectral signature at the position of this given pixel). Hence, a unique foreground/background spectrum is constructed for each of the 2\,378 points (see below) although neighbor pixels have [as it should be] foreground/background material with very similar spectral signatures (i.e., roughly identical linear coefficients). 

The subtraction of the foreground/background emission obviously had a certain incidence on the S/N of the resulting profiles, the peak intensity of all five lines being reduced in the subtracting process. Still following all conditions listed in $\S$~4.2.2, 2\,229 emission-line profiles out of 2\,378 were said to ``survive'' the procedure. Panels (a) and (b) of Figures 16 present the log$_{10}\left[\frac{\textnormal{I}(\textnormal{H}\alpha)}{\textnormal{I}([\textnormal{S}\,\textsc{ii}])}\right]$ vs. log$_{10}\left[\frac{\textnormal{I}(\textnormal{H}\alpha)}{\textnormal{I}([\textnormal{N}\,\textsc{ii}])}\right]$ diagnostic diagram for these 2\,229 points respectively before and after the subtraction of the foreground/background material. Figure 16\textit{a} is obviously a subset of Figure 12\textit{a} and uses the same symbol definitions (see $\S$~4.2.5). Note that each point preserves its original symbol, shape (cross/circle) and color(blue/red), from Figure 16\textit{a} to 16\textit{b}. 

In Figure 16\textit{a}, 2\,109 are displayed as blue crosses while 120 are symbolized as red filled circles, being located close but outside the lower-left corner of the ``H\,\textsc{ii} regions'' area. Note that 134 points were displayed as red circles in Figure 12\textit{a} (see $\S$~4.2.5): 14 of those points were not considered here, either not associated with the central structure or have simply failed to fulfill the conditions of $\S$~4.2.2 following the subtracting process discussed above.

In Figure 16\textit{b}, a tail precisely directed toward the ``SNRs'' area appears as numerous points migrate out of the ``H\,\textsc{ii} regions'' area. Recent surveys (e.g., \citealt{Rie2006,Fre2010}) have shown that the boundaries of the different areas circumscribed in the diagnostic diagrams used here could be slightly modified with respect to the first work carried on by \citet{Sab1977}. It appears that the lower and upper limits, in the parameter space, for each regime are still uncertain to this day. We therefore chose to arbitrarily redefine (or expand) the zone of shock-excitation for each relation investigated. For example, in the log$_{10}\left[\frac{\textnormal{I}(\textnormal{H}\alpha)}{\textnormal{I}([\textnormal{S}\,\textsc{ii}])}\right]$ vs. log$_{10}\left[\frac{\textnormal{I}(\textnormal{H}\alpha)}{\textnormal{I}([\textnormal{N}\,\textsc{ii}])}\right]$ relation, the hatched area in the lower-left corner of Figure 16\textit{b} was, from now on, used to separate gas columns (i.e., pixels) likely enclosing post-shocked material from those for which photoionization effects appear to primarily dominate. Now, 509 points out of 2\,229 suggest evidence for potential shock excitation in the direct vicinity of the Melotte 15 cluster. The tail is formed of both blue crosses and red circles indicating that Figure 16\textit{a} alone would have not be sufficient in order to precisely pinpoint those sections of the central structure revealing (or hiding) a well-defined signature of ionization by shocks. Panel (a) of Figure 17 provides the post-subtraction spatial distribution of all points found in Figure 16\textit{b}. Points found in the hatched area of the lower-left portion of the diagram are colored in red. Others, still associated to the ``H\,\textsc{ii} regions'' area hence suggesting photoionization effects, are in blue.

\begin{figure*}
  \begin{center}
    \leavevmode
    \scalebox{0.8}{\includegraphics[scale=1.00]{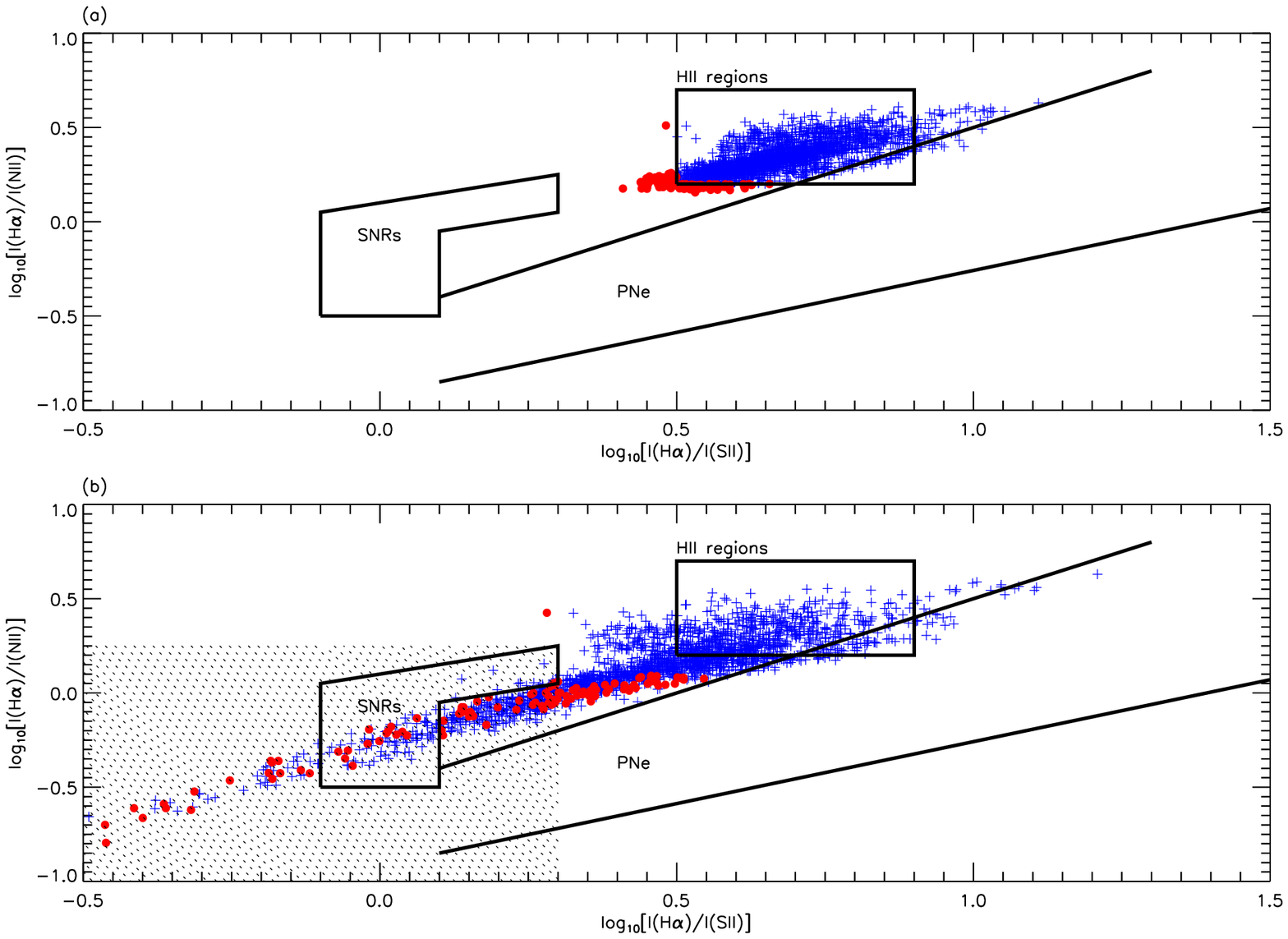}}
  \end{center}
  \caption{Panel (a): Diagnostic diagram of the log$_{10}\left[\frac{\textnormal{I}(\textnormal{H}\alpha)}{\textnormal{I}([\textnormal{S}\,\textsc{ii}])}\right]$ vs. log$_{10}\left[\frac{\textnormal{I}(\textnormal{H}\alpha)}{\textnormal{I}([\textnormal{N}\,\textsc{ii}])}\right]$ relation for points strictly associated to the bright, central structure and prior to the subtraction of the foreground/background material (see text). Definitions for blue crosses and red filled circles are the same as in Figure 12\textit{a}. Panel (b): Diagnostic diagram of the log$_{10}\left[\frac{\textnormal{I}(\textnormal{H}\alpha)}{\textnormal{I}([\textnormal{S}\,\textsc{ii}])}\right]$ vs. log$_{10}\left[\frac{\textnormal{I}(\textnormal{H}\alpha)}{\textnormal{I}([\textnormal{N}\,\textsc{ii}])}\right]$ relation for points strictly associated to the bright, central structure and following the subtraction of the foreground/background material (see text). Each point kept its original symbol used in Panel (a). The hatched area roughly indicates the zone, in the parameter space, where already cataloged shock-dominated objects can be found.}
\end{figure*}

Similar exercises were carried on using the corresponding subsets for Figures 13\textit{a} and 14\textit{a} although we will simply provide the results here. The zone of shock excitation was, respectively for both figures, redefined by log$_{10}\left[\frac{\textnormal{I}(\textnormal{H}\alpha)}{\textnormal{I}([\textnormal{S}\,\textsc{ii}])}\right]$\,$<$\,0.3 and log$_{10}\left[\frac{\textnormal{I}(\textnormal{H}\alpha)}{\textnormal{I}([\textnormal{N}\,\textsc{ii}])}\right]$\,$<$\,0.2, and also bordered by the two long-dashed lines (i.e., lower and upper limits on $\frac{[\textnormal{S}\,\textsc{ii}]\,\lambda6716}{[\textnormal{S}\,\textsc{ii}]\,\lambda6731}$).

For the log$_{10}\left[\frac{\textnormal{I}(\textnormal{H}\alpha)}{\textnormal{I}([\textnormal{S}\,\textsc{ii}])}\right]$ vs. $\frac{[\textnormal{S}\,\textsc{ii}]\,\lambda6716}{[\textnormal{S}\,\textsc{ii}]\,\lambda6731}$ diagram, while only 54 out of 2\,229 points were being found initially outside the ``H\,\textsc{ii} regions'' area prior to the subtraction of the foreground/background material, 509 points have shown log$_{10}\left[\frac{\textnormal{I}(\textnormal{H}\alpha)}{\textnormal{I}([\textnormal{S}\,\textsc{ii}])}\right]$\,$<$\,0.3 afterwards. Again, as in Figure 16\textit{b}, evidences for shocks appear for both blue crosses and red filled circles. Panel (b) of Figure 17 gives the spatial distribution of the post-subtraction points. One clearly sees the perfect correspondence between Panels (a) and (b) i.e., using two distinct diagrams the same points, that suggest shock excitation, are extracted.

For the log$_{10}\left[\frac{\textnormal{I}(\textnormal{H}\alpha)}{\textnormal{I}([\textnormal{N}\,\textsc{ii}])}\right]$ vs. $\frac{[\textnormal{S}\,\textsc{ii}]\,\lambda6716}{[\textnormal{S}\,\textsc{ii}]\,\lambda6731}$ relation, the situation is more complex as the ``SNRs'' and ``H\,\textsc{ii} regions'' areas overlap in the diagnostic tools (see Figure 14\textit{a}). The removal of the foreground/background material has led to 1\,419 points displaying log$_{10}\left[\frac{\textnormal{I}(\textnormal{H}\alpha)}{\textnormal{I}([\textnormal{N}\,\textsc{ii}])}\right]$ values below 0.2 although roughly half of these remain very close to the ``H\,\textsc{ii} regions'' area. For the log$_{10}\left[\frac{\textnormal{I}(\textnormal{H}\alpha)}{\textnormal{I}([\textnormal{N}\,\textsc{ii}])}\right]$ vs. $\frac{[\textnormal{S}\,\textsc{ii}]\,\lambda6716}{[\textnormal{S}\,\textsc{ii}]\,\lambda6731}$ relation, Panel (c) of Figure 17 gives the spatial distribution of the post-subtraction points using the same color code as in Panels (a) and (b).

Combining the post-subtraction results for all three diagrams allows us to spatially identify those sections on the plane of the sky that suggest shock excitation. Panel (d) of Figure 17 combines all results found in the first three panels. Red dots that can be found in all three post-subtraction diagnostic diagrams are again colored in red. These points targeted regions with high probability of shock excitation in central IC\,1805. Those red dots that can only be found in Panel (c) are colored in green in Panel (d). These can be referred as interesting candidates for shock excitation although 2 out of 3 diagrams have failed to identify them. Blue dots in all three post-subtraction diagnostic diagrams are colored in blue and identify the zones, mostly located near the northern tip of the bright ionized feature, where photoionization clearly dominates.

\begin{figure*}
  \begin{center}
    \leavevmode
    \scalebox{0.8}{\includegraphics[angle=90,scale=0.85]{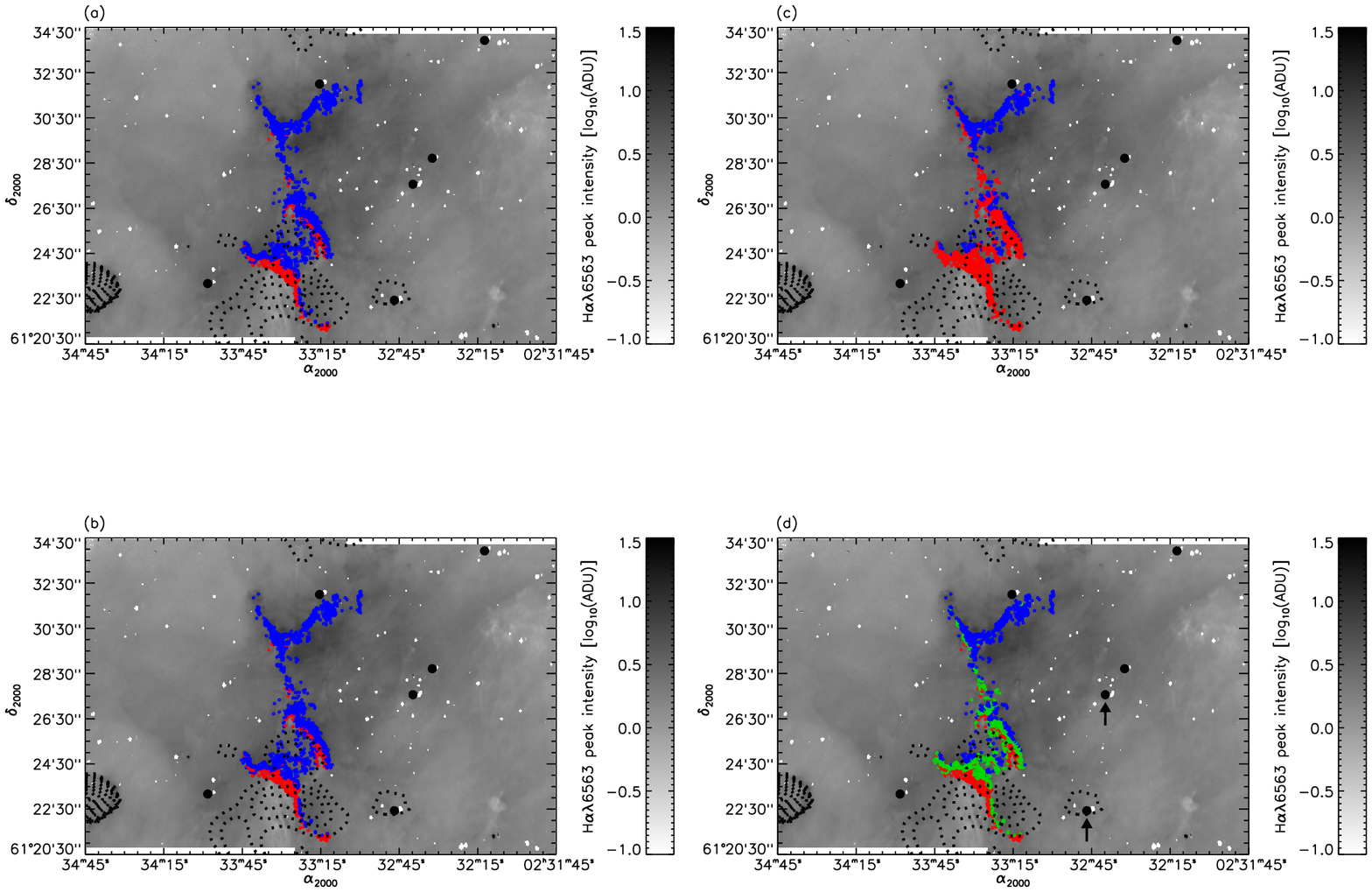}}
  \end{center}
  \caption{Peak-intensity maps of the H{$\alpha$}\,$\lambda$6563 $\mbox{\AA}$ ionic transition in IC\,1805 (see caption of Figure 4). Panel (a): Obtained from the post-subtraction log$_{10}\left[\frac{\textnormal{I}(\textnormal{H}\alpha)}{\textnormal{I}([\textnormal{S}\,\textsc{ii}])}\right]$ vs. log$_{10}\left[\frac{\textnormal{I}(\textnormal{H}\alpha)}{\textnormal{I}([\textnormal{N}\,\textsc{ii}])}\right]$ relation, the specific lines-of-sight, associated to the bright, central structure and suggesting evidences for shock excitation, are represented in red. Blue dots indicate the position where the ionized material is most likely dominated by photoionization effects. Panel (b): Same as Panel (a) for the log$_{10}\left[\frac{\textnormal{I}(\textnormal{H}\alpha)}{\textnormal{I}([\textnormal{S}\,\textsc{ii}])}\right]$ vs. $\frac{[\textnormal{S}\,\textsc{ii}]\,\lambda6716}{[\textnormal{S}\,\textsc{ii}]\,\lambda6731}$ line ratio relation. Panel (c): Same as Panel (a) for the log$_{10}\left[\frac{\textnormal{I}(\textnormal{H}\alpha)}{\textnormal{I}([\textnormal{N}\,\textsc{ii}])}\right]$ vs. $\frac{[\textnormal{S}\,\textsc{ii}]\,\lambda6716}{[\textnormal{S}\,\textsc{ii}]\,\lambda6731}$ line ratio relation. Panel (d): Combination of Panels (a) to (c). Red dots present in all first three panels are again colored in red. Dots colored in red in Panel (c) but blue in Panels (a) and (b) are colored in green. Blue dots present in all first three panels are again colored in blue. The two most massive O-stars, in the Melotte 15 cluster, are identified by the black arrows (see text).}
\end{figure*}

\begin{figure*}
  \begin{center}
    \leavevmode
    \scalebox{0.8}{\includegraphics[scale=1.00]{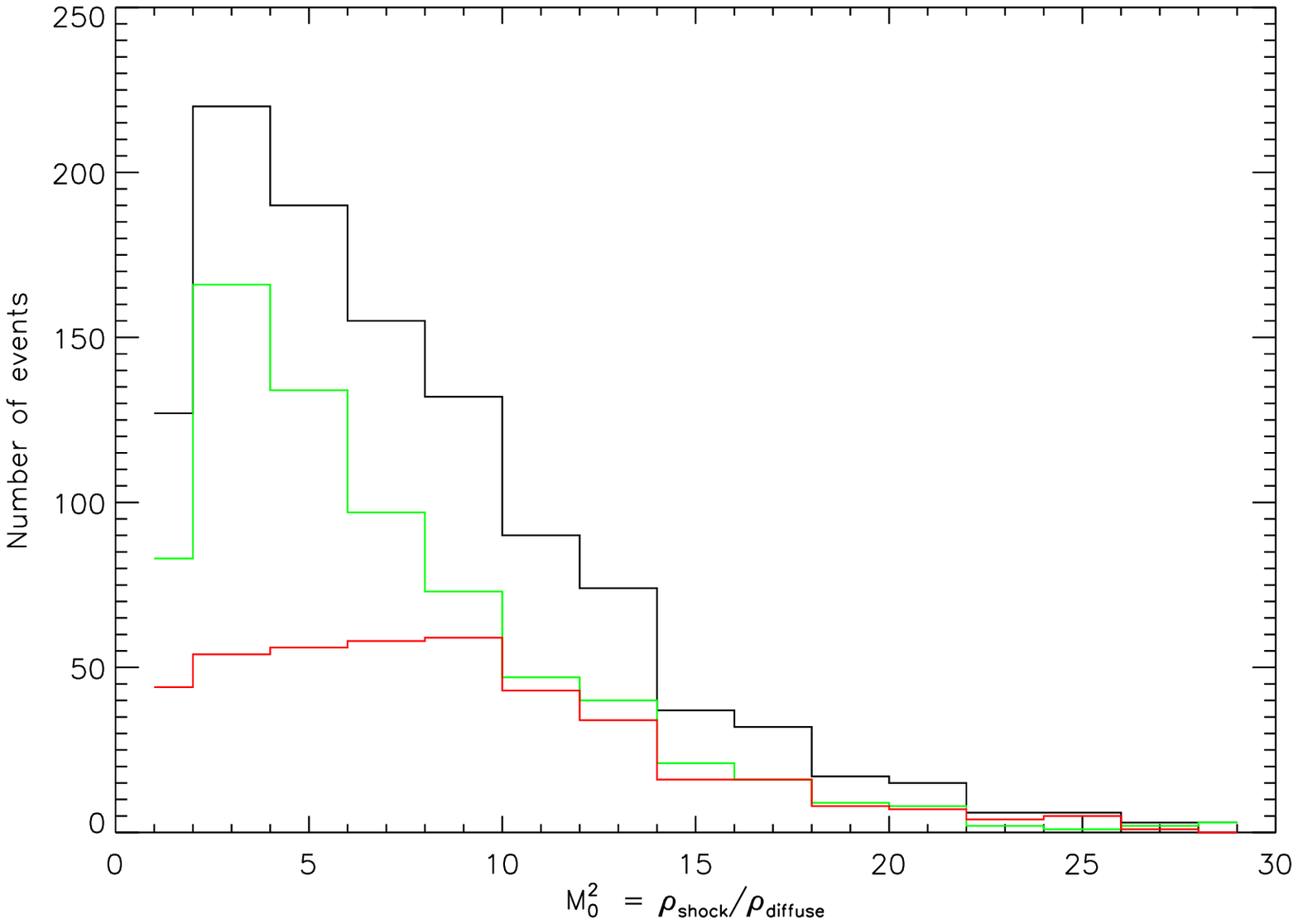}}
  \end{center}
  \caption{Histogram of the compression factor for points, associated to the central structure, revealing post-subtraction evidences for shock excitation (red and green dots in Figure 17\textit{d}). The parameter $\rho_{shock}$/$\rho_{diffuse}$ corresponds, pixel-to-pixel, to the ratio between the electron density computed from the $\frac{[\textnormal{S}\,\textsc{ii}]\,\lambda6716}{[\textnormal{S}\,\textsc{ii}]\,\lambda6731}$ line ratio following the subtraction of the foreground/background material and the electron density of the diffuse component retrieved from the foreground/background spectrum (see text). The green and red distributions are subsets of the black histogram and correspond respectively to the sets of green and red dots found in Figure 17\textit{d}.}
\end{figure*}

Shock-dominated material seems to prevail along the southern portion of the central structure, directly exposed to the stellar winds and currently eroded by the ionizing flux of the two most massive stars identified in Melotte 15 (identified by black arrows in Figure 17\textit{d}). These are cataloged by \citet{Mas1995} as HB\,15558 and HB\,15570, two evolved O4III(f) ($\alpha_{2000}$\,=\,02$^{\textnormal{h}}$32$^{\textnormal{m}}$43$^{\textnormal{s}}$, $\delta_{2000}$\,=\,61$\arcdeg$27$\arcmin$22$\arcsec$) and O4If ($\alpha_{2000}$\,=\,02$^{\textnormal{h}}$32$^{\textnormal{m}}$50$^{\textnormal{s}}$, $\delta_{2000}$\,=\,61$\arcdeg$22$\arcmin$42$\arcsec$) stars respectively. Ionization by shocks clearly dominates the bright ionized gas at the periphery of the weak molecular material found in the southernmost portion of our FOV. Our results reveal that shock excitation appears to be correlated with bright ionized material, suggesting high electron densities if we assume the nebula to be optically-thin to the ionic light. This would then be easily correlated to the compressive nature of the hypothetical shocks.

Extracting the electron densities $\rho$ for all 1\,419 points (rather in red or green in Figure 17\textit{d}) yields the blue-line histogram of Figure 10\textit{a}. Its mean is estimated at 125$^{+\,70}_{-\,50}$ cm$^{-3}$, slightly below the mean value previously derived for the whole sample of 3\,057 emission-line profiles retained for this work (see $\S$~4.2.3). This hardly makes sense considering the compression expected from shock waves. However, keeping in mind that the subtraction of, what we referred to as, the foreground/background material was required in order to spectrally detect evidences for shock excitation in IC\,1805, it seems logical that the same procedure should be applied in the extraction of the electron densities specifically associated to the shocked, ionized component. Using the 1\,419 post-subtraction emission-line profiles that showed evidences for shock excitation, the corresponding electron densities ($\equiv$\,$\rho_{shock}$) were computed following the method described in $\S$~4.2.3. The red histogram, in Figure 10\textit{a}, reveals the distribution obtained for $\rho_{shock}$. The mean electron density along the line-of-sight at the position of shocks is estimated at 175 cm$^{-3}$, a mean roughly 20 cm$^{-3}$ greater than what was found for the black histogram in Figure 10\textit{a}. This suggests, as expected, that the foreground/background material probably occupies, near Melotte 15, a large volume filled with a very diffuse ionized component. This tenuous material undoubtedly contributes to attenuate the ionic emission of high-density condensations along the line-of-sight.

Using the 1\,419 spectra of the foreground/background emission, we extracted to corresponding densities ($\equiv$\,$\rho_{diffuse}$). As expected, this tenuous component has a mean electron density, along the line-of-sight, of 25 cm$^{-3}$. The distribution shows a narrow width of roughly 15 cm$^{-3}$ which reflects the homogeneity (in density) of the foreground/background material over the entire spatial extent of the central, ionized structure. Figure 18 shows the distribution of the pixel-to-pixel compression factor i.e., $\rho_{shock}$/$\rho_{diffuse}$. All remaining 1\,419 points formed the black histogram. The mean compression is estimated at 8.5 although the histogram is clearly dominated by values of $\rho_{shock}$/$\rho_{diffuse}$ below 5. As expected, the compression factor is always greater than 1. Sub-distributions, in green and red, gives the compression factor respectively for the green and red dots obtained in Figure 17\textit{d}. The green histogram, formed of points suggesting possible shock excitation, is very similar statistically to the black distribution with a mean of 7.5. The red histogram, this time formed of points with a high probability of shock excitation, is slightly different, peaking at a greater $\rho_{shock}$/$\rho_{diffuse}$ value. Its mean is estimated slightly above 10 i.e., those points (red in Figure 17\textit{d}) where, we believe, shocks may certainly dominate seem to be characterized by greater compression effects with respect to points (green in Figure 17\textit{d}) where the presence of shocks appears less obvious.

For abiadatic jump conditions, Rankine-Hugoniot relations indicate a compression factor of precisely 4 between pre- and post-shocked gas \citep[Chapter 12]{Hen1999}. The relatively large widths of the distributions displayed in Figure 18 indicate that isothermal shocks may dominate. Also, the fact that forbidden lines such as $[$N\,\textsc{ii}$]$ and $[$S\,\textsc{ii}$]$ are detected favors radiative cooling and therefore non-adiabatic effects [although these lines most likely originate from ionized material trapped in the cooling, shocked outer shell and therefore may not be related to the wind-blown bubble's current state of expansion i.e., energy- or momentum-driven]. Theoretically, for an isothermal shock \citep[Chapter 6]{Dys1980}, the compression factor corresponds to the square of the shock's Mach number, $M_{0}$ (see the abscissa in Figure 18). Therefore, for a speed of sound of 10 km s$^{-1}$ in the H$^{+}$ medium (in agreement with our choice of 7\,400 K for the electron temperature in IC\,1805; see $\S$~4.2.3), shocks with typical velocities between 15 and 30 km s$^{-1}$ (i.e., $M_{0}^2$\,=\,2$-$10) may have led to shock excitation near Melotte 15. This is in agreement with typical expansion velocities found for wind-blown bubbles (see $\S$~1). Shock velocities up to 50 km s$^{-1}$ (i.e., $M_{0}^2$\,$\sim$\,25) are also found although such dynamics is certainly peculiar. Note, however, that the compression measured here could be an upper limit since photoeroded gas located in the molecular cloud's envelope, rather than the diffuse foreground/background component, may act as pre-shocked material. The cloud's envelope is expected to be, at least, a few tens of particles per cm$^{3}$ denser than typical values found for $\rho_{diffuse}$ in this work \citep{Rat2009}.

Assuming a coplanar geometry, roughly 3$-$4 pc separate the most massive star of the Melotte 15 cluster and the central, ionized structure. Using the standard model for expanding wind-blown bubbles \citep{Wea1977}, the distance reached by a given bubble's post-shocked shell can be estimated from the mechanical luminosity of the winds ($L_{w}$\,=\,$\frac{1}{2}$$\dot{M}_{w}$$v_{\infty}^2$), the density of the pre-shocked medium ($\rho_{diffuse}$) and the total time during which winds have been blown by the central star ($t_{w}$). Given $t_{w}$\,=\,2.5 Myr, the age of the Melotte 15 cluster (see $\S$~2), we expect that two massive giant and supergiant O4 stars have left just very recently the main-sequence branch. Therefore, for the calculations to follow, these were treated as O4V stars. From different studies proposing typical mass-loss rates and terminal velocities with respect to spectral types \citep{Pri1990,Lam1999,Mar2004}, $\dot{M}_{w}$ and $v_{\infty}$ were respectively estimated at 5\,$\times$\,10$^{-6}$ M$_{\bigodot}$~yr$^{-1}$ and 3\,000 km s$^{-1}$ for standard O4V stars. Given $\rho_{diffuse}$\,$\sim$\,25 cm$^{-3}$ (see above), nebular material found within 15$-$20 pc of Melotte 15 is likely to have been previously disrupted by stellar winds emanating from the current cluster. This was calculated assuming radiative (non-adiabatic) shocks \citep[Equation 13.5]{Loz1992}. Given that interstellar shocks are usually adiabatic at the earliest times of the expansion, the non-radiative model yields a distance of roughly 40 pc in the vicinity of Melotte 15 \citep[Equation 13.9]{Loz1992}. Both distances exceed the extent of our FOV (15\,$\times$\,9 pc$^{2}$) and support the fact that shocks, attributed to the current star cluster, have had sufficient time to reach the bright structure in central IC\,1805.  

\subsubsection{Shock excitation in other zones of the nebula}

Besides the bright, central structure observed near the most massive stars of the Melotte 15 cluster, our FOV mostly reveals diffuse ionized gas, partially obscured by interstellar dust. However, two structures are nonetheless detected, particularly well-defined in H{$\alpha$} (see $\S$~4.1). In order to investigate the presence of shock excitation in other portions of our FOV, a similar approach, to what was carried on in $\S$~5.2.1, is used here on these two features. 

\paragraph{South-east CO fragment.}

Panel (a) of Figure 19 reveals the shape of the ionization front observed at the periphery of the CO fragment found in the south-east portion of our FOV (see Figure 8). As in Figures 3 to 9, North is up and East is left. The molecular feature is well-defined in the millimeter regime and its western side shows signs of erosion by the nearby star cluster. 

\begin{figure*}
  \begin{center}
    \leavevmode
    \scalebox{0.8}{\includegraphics[angle=90,scale=0.95]{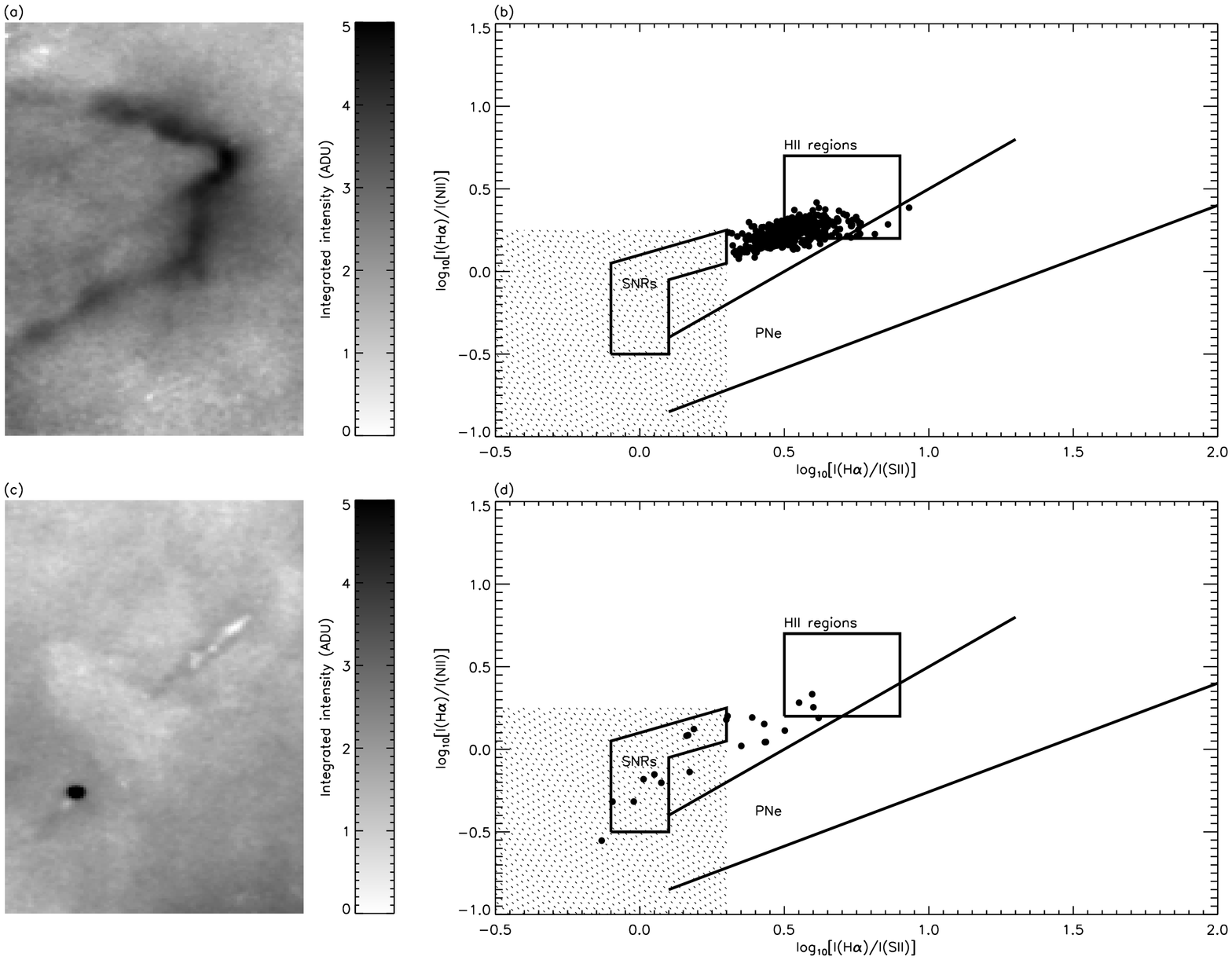}}
  \end{center}
  \caption{Investigating the presence of shock excitation in other portions of IC\,1805. Panel (a): Integrated intensity of the H{$\alpha$} ionic transition near the CO fragment, detected by the FCRAO survey, and located in the south-east portion of our FOV (see $\S$~4.1). Panel (b): Diagnostic diagram of the log$_{10}\left[\frac{\textnormal{I}(\textnormal{H}\alpha)}{\textnormal{I}([\textnormal{S}\,\textsc{ii}])}\right]$ vs. log$_{10}\left[\frac{\textnormal{I}(\textnormal{H}\alpha)}{\textnormal{I}([\textnormal{N}\,\textsc{ii}])}\right]$ relation for points associated to the bright ionization front displayed in Panel (a). The diagram is obtained following the subtraction of the foreground/background material (see text). Panel (c): Integrated intensity of the H{$\alpha$} ionic transition associated to a cigar-like structure pointing toward a nearby star with strong H{$\alpha$} emission. Panel (d): Diagnostic diagram of the log$_{10}\left[\frac{\textnormal{I}(\textnormal{H}\alpha)}{\textnormal{I}([\textnormal{S}\,\textsc{ii}])}\right]$ vs. log$_{10}\left[\frac{\textnormal{I}(\textnormal{H}\alpha)}{\textnormal{I}([\textnormal{N}\,\textsc{ii}])}\right]$ relation for points associated to the cigar-like structure displayed in Panel (c). The diagram is obtained following the subtraction of the foreground/background material (see text). For Panels (a) and (c), both maps were processed using the spectral collapse of the initial mosaicked cube in the channel interval of the H{$\alpha$}\,$\lambda$6563 $\mbox{\AA}$ ionic line.}
\end{figure*} 

The ionized gas associated to the ionization front was first circumscribed while weaker areas near the bright H$^{+}$ feature were selected and used to extract the foreground/background spectrum. In order to obtain a post-subtraction sample containing a sufficient number of emission-line profiles, only the first condition of $\S$~4.2.2 was used. Therefore, results presented here should be cautiously interpreted. 

Prior to the subtraction, all points were initially found inside the ``H\,\textsc{ii} regions'' area of the log$_{10}\left[\frac{\textnormal{I}(\textnormal{H}\alpha)}{\textnormal{I}([\textnormal{S}\,\textsc{ii}])}\right]$ vs. log$_{10}\left[\frac{\textnormal{I}(\textnormal{H}\alpha)}{\textnormal{I}([\textnormal{N}\,\textsc{ii}])}\right]$ diagnostic diagram and black filled circles were used as symbols. Panel (b) displays the diagram for post-subtraction spectra only. These results reveal, as in $\S$~5.2.1, that points are displaced toward the shock-dominated ``SNRs'' area although the tail is much less developed when compared to Figure 16\textit{b}. No post-subtraction point actually enters the hatched area defined earlier. 

Shock excitation hence appears to have played a much more minor role in the ionization of this cloud's envelope when compared to the central structure of our FOV. The apparent shape of the ionization front indicates that the ionizing sources are located behind the molecular cloud (see $\S$~4.1). Using kinematical information retrieved from the photoevaporated material in the vicinity of the cloud, \citet{Lag2009a} suggested that an appreciable distance could exist between the CO fragment and the Melotte 15 cluster (see the authors' Figure 11 and their associated Flow G). Hence, large distances could signify here that stellar-wind shocks may have been partially dissipated before they could reach the molecular cloud. Shock excitation is therefore likely measurable only within a certain distance of the shock sources i.e., stars with strong stellar winds in our case. Obviously, the stronger the shocks are, the greater this corresponding distance is. We reiterate that only the first 15 to 40 pc in the vicinity of Melotte 15 may have been disrupted by stellar winds (see $\S$~5.2.1). At this point, the strongest wind shocks (fuelled up by the O4 stars) would have receded to the subsonic regime with velocities between 4 and 9 km s$^{-1}$ depending on the model used (radiative or non-radiative respectively). If, by any chance, the molecular fragment found near the south-eastern boundary of our FOV is located relatively far from the ionizing sources ($\gtrsim$\,15 pc), this could explain the apparent absence of strong, compressive shocks at its periphery.  

\paragraph{Cigar-like structure.}

Panel (c) of Figure 19 shows the very tenuous, but well-defined, ionized counterpart of a cigar-like structure pointing toward a nearby star with strong H{$\alpha$} emission line. Again, North is up and East is left. The star is particularly visible on the left-hand side of Figure 8 and has been cataloged by \citet{Koh1999} as HBH{$\alpha$}\,6211-05 with a V-band magnitude of 14.6 and no known spectral type. Its coordinates are ($\alpha_{2000}$\,=\,02$^{\textnormal{h}}$34$^{\textnormal{m}}$10$^{\textnormal{s}}$.06, $\delta_{2000}$\,=\,61$\arcdeg$24$\arcmin$35$\arcsec$.7). The elongated feature was briefly introduced in $\S$~4.1.

Emission in H{$\alpha$} and $[$N\,\textsc{ii}$]$ is detected on the outskirts of the cigar-like feature while its center appears mostly gas depleted. This overall scheme is particularly similar to what is expected from elephant trunks in H\,\textsc{ii} regions (e.g., \citealt{Car2003}). Shadowing effects, caused by a dense neutral globule located in a radiation field, allow the warm gas behind it to recombine which could explain the absence of ionized material at the center of the elongated structure. The bright ionized rims are formed of photoevaporated material created by the erosion of the neutral globule. These flows move away from the ionizing star creating the cigar-like shape.

Both the FCRAO CO(1-0) survey and the Canadian Galactic Plane Survey at 21 cm \citep{Nor1997} reveal no indication for either molecular or atomic gas associated to the elongated feature. Theoretically, H$_{2}$ gas is expected to constitute the main component of the eroded globule while H\,\textsc{i} material should result from recombinations in the tail. The whole structure, however, has relatively small angular dimensions (45$\arcsec$\,$\times$\,7$\arcsec$) which suggests that emission from the neutral gas could be beam diluted in the low-resolution radio observations. 

Theory has suggested that shocks will develop in elephant trunks \citep{Mac2010,Rag2010} although, in first approximation, the log$_{10}\left[\frac{\textnormal{I}(\textnormal{H}\alpha)}{\textnormal{I}([\textnormal{S}\,\textsc{ii}])}\right]$ vs. log$_{10}\left[\frac{\textnormal{I}(\textnormal{H}\alpha)}{\textnormal{I}([\textnormal{N}\,\textsc{ii}])}\right]$ diagram has shown nothing particular using the first set of Gaussian fits i.e., all points associated to the elongated structure are well-confined into the ``H\,\textsc{ii} regions'' area. However, subtracting the foreground/background material yields the diagram displayed in Panel (d) of Figure 19. None of the conditions listed in $\S$~4.2.2 were considered. Only a S/N greater than 3 (sufficiently high to confirm a physical detection; see $\S$~4.2.2) was here required. Therefore, the results are highly questionable although worth mentioning. The ionized material at the center of the cigar-like feature being very tenuous (or simply non-existent; see above), all points filling the diagram are found along the external bright rims. Panel (d) reveals that the ``SNRs'' area of shock-dominated material is highly favored, in agreement with the prediction of shock development proposed by the theory.

The particular shape of the elongated structure presented in Panel (c) could be easily mistaken for jets attributed to Herbig-Haro (HH) objects. However, Figure 4\textit{a} of \citet{Fre2010} indicates the peculiar low intensities of the $[$N\,\textsc{ii}$]$ lines in HH objects. This leads to a position, for HH objects in the log$_{10}\left[\frac{\textnormal{I}(\textnormal{H}\alpha)}{\textnormal{I}([\textnormal{S}\,\textsc{ii}])}\right]$ vs. log$_{10}\left[\frac{\textnormal{I}(\textnormal{H}\alpha)}{\textnormal{I}([\textnormal{N}\,\textsc{ii}])}\right]$ diagram, clearly above the ``SNRs'' area. This said, the diagram of Panel (d) most likely confirms that the cigar-like feature is not part of an HH object in IC\,1805. Rather, the elephant-trunk nature, suggested here, appears very plausible.

\section{Conclusion}

The use of the imaging Fourier transform spectrometer SpIOMM allowed us to obtain series of emission-line profiles of the optical gas in the brightest, central portions of the Galactic IC\,1805 H\,\textsc{ii} region. The bandwidth used at data acquisition allowed the simultaneous observations of the H{$\alpha$}\,$\lambda$6563 $\mbox{\AA}$, $[$N\,\textsc{ii}$]$\,$\lambda$$\lambda$6548, 6584 $\mbox{\AA}$, and $[$S\,\textsc{ii}$]$\,$\lambda$$\lambda$6716, 6731 $\mbox{\AA}$ ionic lines (see $\S$~3). 

The main goal of this work was to investigate on the presence of supersonic shock waves attributed to stellar winds in the vicinity of Melotte 15, the current star cluster actually fueling up the expansion of the IC\,1805 nebula. Literature has long suggested that a kinematical detection of stellar winds, in H\,\textsc{ii} region, might represent a difficult task since the typical expansion velocity of shocked shells could be commonly confused with other dynamical processes revealing similar kinematical behaviors (see $\S$~1). On the other hand, specific line ratios retrieved from the optical gas may indicate if the presence of ionized material is attributed to standard photoionization or to shock excitation. These line ratios therefore provide a non-kinematical tool for identifying shocks in the nebular volume. Our results are summarized as follow:

\begin{enumerate}
\renewcommand{\labelenumi}{\Roman{enumi}.}
\item The $\frac{[\textnormal{N}\,\textsc{ii}]\,\lambda6584}{[\textnormal{N}\,\textsc{ii}]\,\lambda6548}$ line ratio genuinely deviates from the theoretical value of 3 (see $\S$~5.1). Values varying between 2.5 and 4 were commonly found. The distribution of the density measurements has not allowed to demonstrate that the $[$N\,\textsc{ii}$]$\,$\lambda$6548 $\mbox{\AA}$ line could be affected by collisional de-excitations, hence reducing its peak intensity. This scenario remains nonetheless plausible and could be verified if densities, in the N$^{+}$ volume, could be measured precisely (see $\S$~5.1). Densities extracted from the $\frac{[\textnormal{S}\,\textsc{ii}]\,\lambda6716}{[\textnormal{S}\,\textsc{ii}]\,\lambda6731}$ ratio may not perfectly reflect the conditions prevailing in the N$^{+}$ volume if both nitrogen and sulfur are not co-spatial in central IC\,1805.
\item The diagnostic diagrams, introduced by \citet{Sab1977}, indicate, in first approximation, that photoionization most likely dominate in IC\,1805 (see $\S$~4.2.5). This initially holds even for the densest, most emissive structures in our field-of-view, directly exposed to the radiation fields and stellar winds of the nearby, most massive stars (see $\S$~5.2).
\item Evidences for shock excitation appear only following the subtraction of the diffuse foreground/background material. This component likely occupies a very large fraction of the nebular volume in IC\,1805; a volume sufficiently large that, even though being tenuous in nature, the foreground/background material strongly dilutes the signal emanating from shocked condensations along the line-of-sight. Its mean density is estimated at 25 cm$^{-3}$ (see $\S$~5.2.1).
\item Oriented on a south-north axis and surrounded by numerous O-stars, a bright, large ionized feature occupies the central area of our field-of-view. The last, tenuous fragments of an old molecular cloud can be found near its southern portion. Shocks may have contributed to ionize material found in the direct vicinity of the molecular clump while the northern parts of the ionized feature, deprived of molecular emission, appears to be largely dominated by photoionization (see $\S$~5.2.1).
\item The shock-excited ionized gas has a mean density of 175 cm$^{-3}$ and the compression factor, between pre- and post-shocked gas, is typically between 2 and 10. For isothermal shocks, this suggests shock velocities between 15 and 30 km s$^{-1}$, in agreement with models describing the expansion of wind-blown bubbles (see $\S$~5.2.1). Geometrically speaking, given the apparent proximity between the central, ionized structure and the most massive stars of Melotte 15, winds seem to have had sufficient time to reach the structure within a timescale corresponding to the age of the star cluster (see $\S$~5.2.1). This gives credence to our assumption that shock excitation can be found in the south-central portions of our field-of-view.
\item Points identified with a high probability of shock excitation reveal compression factors typically greater than those points where shocks may be present although less certain (see $\S$~5.2.1).
\item Shocks did not seem to have played a major role in the ionization of a molecular cloud's envelope located in the south-eastern portion of our field-of-view. This most likely results from the molecular fragment being located too far from the ionizing sources so that (1) shocks induced by stellar winds have not reached yet the cloud or (2) shocks have reached the cloud with velocities too low to initiate shock excitation (see $\S$~5.2.2.1).
\item Shock development was clearly detected on the outskirts of an apparently weak, but well-defined elephant trunk located in the eastern portion of our field-of-view (see $\S$~5.2.2.2). This is in agreement with theoretical works developed on such feature typically found in H\,\textsc{ii} regions.
\end{enumerate}

\section*{Acknowledgments}

The authors would like to thank the Natural Sciences and Engineering Research Council of Canada and the Fonds Qu\'eb\'ecois de la Recherche sur la Nature et les Technologies who provided funds for this research project.

D. L. is grateful to M.-A. Miville-Desch\^enes and D. J. Marshall who provided useful \textsc{idl} routines to carry out data reduction. D. L. and L. D. would like to thank B. Malenfant, G. Turcotte and P.-L. L\'evesque for technical support during numerous observing nights at the Observatoire du Mont-M\'egantic and also A.-P. Bernier and M. Charlebois who provided convenient help during data acquisition. The development of SpIOMM is a collaboration between Universit\'e Laval and ABB Bomem.

Finally, we are especially thankful to our referee, Dr. Luis Cuesta, for very relevant suggestions that were used to improve the manuscript.

\label{lastpage}

\end{document}